\documentclass[a4paper,11pt]{article}
\usepackage{eurosym}
\PassOptionsToPackage{hyphens}{url}\usepackage{hyperref}
\usepackage{authblk}
\usepackage[nottoc,notlot,notlof]{tocbibind}
\usepackage{tikz-cd}
\usetikzlibrary{arrows, arrows.meta, cd}
\usepackage[all,cmtip]{xy}
\usepackage{amssymb,amsmath,amsthm,mathrsfs}
\usepackage{graphicx}
\usepackage{mathtools}
\usepackage{amsmath}
\usepackage{amssymb}
\usepackage{amsthm}
\usepackage{amsfonts}
\usepackage{mathrsfs}
\usepackage{enumerate}
\usepackage{graphicx,tikz-cd,pgf}
\usepackage[margin=2.5cm]{geometry}
\usepackage[normalem]{ulem}
\usepackage[T1]{fontenc}
\usepackage[utf8]{inputenc}
\usepackage[english]{babel}
\usepackage{url}
\usepackage{stix}
\usepackage{bigints}

\newtheorem{theorem}{Theorem}[section]

\newtheorem{lemma}[theorem]{Lemma}
\newtheorem{definition}[theorem]{Definition}

\newtheorem{example}[theorem]{Example}
\newtheorem{examples}[theorem]{Examples}
\newtheorem{proposition}[theorem]{Proposition}
\newtheorem{corollary}[theorem]{Corollary}
\newtheorem{remark}[theorem]{Remark}

\newcommand{\dr}[1]{
\widetilde{#1}
}

\newcommand{\cil}{\rho[x,n]}
\newcommand{\vcil}{r(x,n)}
\newcommand{\drcil}{\tilde{\rho}[x,n]}

\DeclareMathOperator{\linearhullsymbol}{\mathtt{span}}
\newcommand{\linearhull}[1]{
\linearhullsymbol(#1)
}

\DeclareMathOperator{\primo}{(\mathbf{I})}
\DeclareMathOperator{\secondo}{(\mathbf{II})}

\DeclareMathOperator{\linealitysymbol}{\mathtt{lineal}}
\newcommand{\lineal}[1]{
\linealitysymbol(#1)
}

\DeclareMathOperator{\suppsymbol}{\mathsf{ccSupp}}
\newcommand{\supp}[1]{
\suppsymbol(#1)
}

\DeclareMathOperator{\tsuppsymbol}{\mathtt{tSupp}}
\newcommand{\tsupp}[1]{
\tsuppsymbol(#1)
}

\DeclareMathOperator{\gcone}{\mathsf{K}}
\DeclareMathOperator{\gconetwo}{\mathsf{B}}
\DeclareMathOperator{\ray}{\mathsf{R}}

\newcommand{\nproj}[3]{
{#3}({#1}|#2)
}

\DeclareMathOperator{\extremalsymbol}{\mathtt{eRays}}
\newcommand{\extremal}[1]{
\extremalsymbol(#1)
}

\DeclareMathOperator{\net}{\mathtt{NG}}

\DeclareMathOperator{\NN}{{\mathbb{N}}}
\DeclareMathOperator{\RR}{{\mathbb{R}}}

\DeclareMathOperator{\basissymbol}{\epsilon}
\newcommand{\basis}[1]{
\basissymbol[x,#1]
}

\DeclareMathOperator{\mercato}{{\mathsf{V}}}
\DeclareMathOperator{\mutuum}{{\mathsf{M}}}

\DeclareMathOperator{\mercatocinque}{\mathsf{CC}}

\DeclareMathOperator{\raggio}{\boldsymbol{\mathsf{R}}}

\newcommand{\rgb}[1]{
\raggio[#1]
}

\newcommand{\ccgb}[1]{
\mercatocinque(#1)
}

\DeclareMathOperator{\elemsymbol}{\mathcal{E}}
\DeclareMathOperator{\elemsymbolcone}{\mathsf{E}}
\newcommand{\elemd}[2]{
\elemsymbolcone^{#2}(#1)
}

\newcommand{\element}[1]{
\elemsymbolcone^{#1}
}

\newcommand{\elem}[1]{
\elemsymbol({#1})
}

\DeclareMathOperator{\loansymbol}{\mathsf{L}}
\DeclareMathOperator{\loans}{\loansymbol}
\DeclareMathOperator{\rloansymbol}{\mathsf{R}}
\DeclareMathOperator{\rloans}{\rloansymbol}
\DeclareMathOperator{\ploansymbol}{\mathsf{T}}
\DeclareMathOperator{\ploans}{\ploansymbol}

\DeclareMathOperator{\ccones}{\boldsymbol{\mathsf{C}}}

\DeclareMathOperator{\arbitsymbol}{\mathcal{A}}
\DeclareMathOperator{\arbit}{\arbitsymbol}

\DeclareMathOperator{\sloansymbol}{\mathsf{S}}
\newcommand{\sloans}[1]{
\sloansymbol({#1})
}

\DeclareMathOperator{\annob}{\mathtt{Y}}
\DeclareMathOperator{\anno}{\boldsymbol{\annob}}

\DeclareMathOperator{\meseb}{\mathtt{M}}
\DeclareMathOperator{\mese}{\boldsymbol{\meseb}}

\DeclareMathOperator{\tempob}{\mathtt{T}}
\DeclareMathOperator{\tempo}{\boldsymbol{\tempob}}

\newcommand{\polinomi}{\mathbb{R}[\tempo]}
\newcommand{\polinomia}{\mathbb{R}[\anno]}

\title{On the Behavior of the Payoff Amounts  
\\ in Simple Interest Loans in Arbitrage-Free Markets}

\author{Fausto Di Biase, Stefano Di Rocco, 
Alessandra Ortolano, Maurizio Parton\\
Fausto Di Biase\\
Dipartimento di Economia,
Universit\`a ``G. d'Annunzio'' di Chieti-Pescara\\
Viale Pindaro 42, 65127 Pescara \\
fausto.dibiase@unich.it\\
Stefano Di Rocco\\
Dipartimento di Economia,
Universit\`a ``G. d'Annunzio'' di Chieti-Pescara\\
Viale Pindaro 42, 65127 Pescara \\stefano.dirocco@studenti.unich.it\\
Alessandra Ortolano\\
Dipartimento Economia, Ingegneria, 
Societ\`a
 e Impresa (DEIM)\\ 
 Universit\`a della Tuscia di Viterbo\\
Via del Paradiso 47, 01100 Viterbo VT\\
alessandra.ortolano@unitus.it\\
Maurizio Parton\\
Dipartimento di Economia, 
Universit\`a ``G. d'Annunzio'' di Chieti-Pescara\\
Viale Pindaro 42, 65127 Pescara \\
maurizio.parton@unich.it}

\begin{document}

\maketitle


\begin{abstract}{
The Consumer Financial Protection Bureau 
defines the notion of \textit{payoff amount} as 
the amount that has to be payed
at a particular time 
in order to completely pay off 
the debt, in case 
the lender intends to pay off the loan early, way before
the last installment is due (CFPB 2020). 
This amount is well-understood for loans 
at compound interest, but much less so when 
simple interest is used. Recently, 
Aretusi and Mari (2018) have proposed a  \textit{formula 
for the payoff amount} for 
 \textit{loans at simple interest}. We assume that the payoff amounts are established contractually at time zero, whence the requirement that no arbitrage may arise this way

The first goal of this paper is to study this new formula
and derive it within a  \textit{model of a loan market} in which loans are bought and sold at simple interest, interest rates change over time, no arbitrage opportunities exist, 
there are no transaction costs or insolvency risks, and the usual symmetry hypothesis holds: all economic agents have equal opportunity to buy or sell the loans available in the loan market. 

The second goal is to show that 
this formula exhibits a  \textit{behaviour} rather different from the one which occurs when compound interest is used. Indeed, we show that, if the installments are constant and 
if the interest rate is greater than a critical value (which depends on the number of installments), then the sequence of the payoff amounts is increasing before a certain critical time, and will start decreasing only thereafter. 
This behavior is at odds with the familiar intuition that comes from compound interest, where the sequence of the payoff amounts immediately starts decreasing
(indeed, several of the formulas that are familiar within 
the context of compound interest do not hold  in this model).
We also show that the critical value is decreasing as a function of the number of installments. 
For two installments, the critical value is equal to the golden section.

The third goal is to 
introduce a more efficient \textit{polynomial notation}, which encodes 
a basic tenet of the subject: 
Each amount of money 
is \textit{embedded 
in a time position} (to wit: The time when it is due). 
The model of a loan market we propose is naturally linked to this new notation.}
\end{abstract}


\paragraph{Acknowledgments.} The first-named author was partially supported by INdAM-GNAMPA and the fourth by INdAM-GNSAGA. 

\section*{Compliance with Ethical Standards and Ethical Statement} 
No competing interests or conflict of interest, neither financial nor non-financial, that are directly or indirectly related to the present work, exist for any of the authors or have existed in the past.  This research does not involve and has not involved Human Participants and/or Animals.

\section{Introduction}
The Consumer Financial Protection Bureau 
defines the notion of \textit{payoff amount} as 
the amount that has to be payed
at a particular time 
in order to completely pay off 
the debt, in case 
the lender intends to pay off the loan early, way before
the last installment is due (CFPB 2020). 
This amount is well-understood for loans 
at compound interest, but much less so when 
simple interest is used. Recently, 
Aretusi and Mari (2018) have proposed a formula 
for the payoff amount for 
 \textit{loans at simple interest}. We assume that the payoff amounts are established contractually at time zero, whence the requirement that no arbitrage may arise this way.

The first goal of this paper is to study this new formula
and derive it within a  \textit{model of a loan market} in which loans are bought and sold at simple interest, interest rates change over time, no arbitrage opportunities exist, 
there are no transaction costs or insolvency risks, and the usual symmetry hypothesis holds: all economic agents have equal opportunity to buy or sell the loans available in the loan market. 

The second goal is to show that 
this formula exhibits a  \textit{behaviour} rather different from the one which occurs when compound interest is used. Indeed, we show that, if the installments are constant and 
if the interest rate is greater than a critical value (which depends on the number of installments), then the sequence of the payoff amounts is increasing before a certain critical time, and will start decreasing only thereafter. 
This behavior is at odds with the familiar intuition that comes from compound interest, where the sequence of the payoff amounts immediately starts decreasing
(indeed, several of the formulas that are familiar within 
the context of compound interest do not hold  in this model).
We also show that the critical value is decreasing as a function of the number of installments. 
For two installments, the critical value is equal to the golden section.

The third goal is to 
introduce a more efficient \textit{polynomial notation}, which encodes 
a basic tenet of the subject: 
Each amount of money 
is \textit{embedded 
in a time position} (to wit: The time when it is due). 
The model of a loan market we propose is naturally linked to this new notation.

\section{Notation for  Financial Operations}\label{s:notation}

In our model of a \textit{loan market}, time is discrete. 
Indeed, the background of this work is the intent to include
in a mathematical model  some of the Italian laws related to  \textit{lending} (which, in this legislation, are seen as distinct from \textit{investment instruments}).
In this  framework, the minimum time interval to be considered is the day (art. 821 Italian Civil Code). 

We introduce a new and, we believe, more efficient \textit{polynomial notation} that encodes 
a basic tenet of the subject, according to which
each amount of money 
is \textit{embedded 
in a time position}, to wit: The time when it is due. 
Our notation is best 
explained by some examples.
The polynomial
\begin{equation}
-100+10\annob+110\annob^2
\label{eq:firstexample}
\end{equation}
(the indeterminate $\anno$ stands for \textit{year}) 
encodes the cash flow associated to a financial operation 
where an economic agent \textit{pays} 100 at time zero, \textit{receives} 10 precisely one year 
later, and finally \textit{receives} 110 precisely two years after time zero (in order to reduce notational clutter, 
we do not indicate the currency). 
By the same token, 
\begin{equation}
-100+120\annob^2
\label{eq:secondexample}
\end{equation}
encodes the cash flow associated to a financial operation 
where an economic agent \textit{pays} 100 at time zero
and \textit{receives} 120 precisely two years 
later. Here follows a general description of our notation. 
\paragraph{Notation and Interpretation.}
A polynomial such as 
$$
\alpha=\sum_{k=0}^{N}\alpha(k) \anno^k
$$
encodes the cash flow of a financial operation with at most $N$ yearly payments after the one at time zero. 
The sign of the coefficient $\alpha(k)$ 
indicates whether the amount is 
due to us (if positive) or if we owe it (if negative). 
The absolute value 
$\vert\alpha(k)\vert$, where $k\geq 1$,  
is called the $k$-th \textbf{installment} associated to $\alpha$
and 
denotes the amount to be payed precisely $k$ years after time zero.
The \textbf{maturity} of $\alpha$ is the degree of $\alpha$:
If $\alpha(N)\not=0$ then we say that $N$ is the \textbf{maturity} of 
$\alpha$.  

\begin{definition}
If $\alpha\in\polinomi$, the \textbf{(ordinary) support} of $\alpha$ is defined as
\begin{equation}
\tsupp{\alpha}\eqdef\{k\in\NN:\alpha(k)\not=0\}
\label{eq:tsupport} 
\end{equation}
\label{d:tsupport}
\end{definition}

\paragraph{Interpretation.} In our setting, the (ordinary) support of 
$\alpha$, when $\alpha$ is seen as a financial operation, identifies 
the positions in time when the payments are due, according to the 
underlying contract. 
Indeed, the parties involved  have the contractual freedom to 
choose when the payments are due (as long as the contract is compatible 
with existing laws). In~\eqref{eq:support} we will introduce a different notion 
of support, related to the notion of extremal ray in a convex cone. In order to avoid 
confusion between the two notions, they are denoted by different symbols.

\begin{remark}
Strictly speaking, 
a financial operation is a legal instrument, of which 
the associated cash flow
 only describes in part the accounting side. 
 However, we will simply 
 speak of \textbf{the financial operation $\alpha\in\polinomia$}.
 \end{remark}

\begin{remark}
The vector space structure of 
$\polinomia$ has a natural bearing on this subject, since the sum of two polynomials is a representation of the total cash flow that results from the superposition of two financial operations, and multiplication by a scalar is a representation of the total cash flow that results from operating with many copies of a given financial operation. 
\end{remark}

\begin{remark}
In concrete applications, 
an amount such as 
\euro\, $10^{-100}$ does not arise and it has to be treated as zero. 
This fact suggests that  
the ring of coefficients contains nilpotent elements (and hence is not equal to  $\RR$).
However,  we will ignore this point, which is, after all, virtually ubiquitous in numerical applications.  
\end{remark}

\begin{examples}
In a similay vein, a polynomial 
$$
\beta=\sum_{k=0}^{N}\beta(k) \mese^k
$$
encodes the cash flow of   a financial operation with 
 \textit{monthly} payments. In general,
the notation 
$$
\gamma=\sum_{k=0}^{N}\gamma(k) \tempo^k
$$
(where  $\tempo$ is a given unit of time)
encodes the cash flow associated to a 
financial operation where the payments $\gamma_k$ are \textit{due} at ``time'' $k$, with respect to the given time unit
$\tempo$. The symbol $\tempo$ 
denotes the given time unit and, at the same time, it is used as the ``indeterminate'' in the space $\polinomi$ of all polynomials 
in $\tempo$. 
In short: The cash flows associated to 
financial operations with periodic payments, occurring at times that are powers 
of\/ $\tempo$, are encoded by elements of the space 
$\polinomi$ of all polynomials 
with real coefficients 
in the indeterminate $\tempo$. 
 \end{examples}

In order to fix ideas, we only consider 
financial operations with \textit{yearly} payments, i.e., 
the case 
$\tempo=\anno$. 
An introduction to these topics can be found in Cacciafesta (2001), 
Levi (1953), Varoli (1983).

\section{The Category of Convex Cones}
The \textit{category} $\ccones$ of \textit{convex cones} has some relevance to this subject. 
We only consider (not necessarily finite dimensional) vector spaces over $\RR$.
The zero vector of a given vector space is usually
 denoted by $0$. 
\begin{definition}
Let $\gcone$ be a nonempty subset of a vector space $V$. 
We say that $\gcone$ 
is  a \textbf{cone} (in $V$) if $\alpha\in \gcone$  and 
$r\geq0$ imply that $r\alpha\in \gcone$.
It is  
\textbf{convex} if 
$\alpha,\beta\in \gcone$ and $t\in(0,1)$ imply that 
$t\alpha+(1-t)\beta \in \gcone$.
It is a 
\textbf{convex cone} if it is a cone and it is convex. 
\end{definition}
In this definition,
the set $\{0\}$, which only contains the zero element of vector space, is a convex cone.
Observe that a subset $\gcone$ 
of a vector space 
is a  
convex cone if and only if 
the following properties hold.
\begin{description}
\item[(cc1)] If $\alpha\in \gcone$  and 
$r>0$, then $r\alpha\in \gcone$.

\item[(cc2)] If $\alpha\in \gcone$
and $\beta\in \gcone$, then $\alpha+\beta\in \gcone$.
\end{description} 
Moreover, these two properties are jointly 
equivalent to the following one:
\begin{description}
\item[(cc)]  If $\alpha\in \gcone$
and $\beta\in \gcone$ and $r,s>0$ then 
$r\alpha+s\beta\in\gcone$.
\end{description}

\paragraph{Interpretation and Motivation.}
These algebraic structures 
arise in a natural way from the fact that 
the sum of two polynomials is a representation of the total cash flow that results from the superposition of two financial operations, and multiplication by a scalar is a representation of the total cash flow that results from acting with many copies 
of a given financial operation.

\paragraph{Morphisms in the Category of Convex Cones.}

Convex cones form the objects of a category, denoted by 
$\ccones$, 
where morphisms are given as follows: If $\gcone$ and $\gconetwo$ are convex cones (in vector spaces $V$ and $V'$) then a 
\textbf{$\ccones$-homomorphism} 
\textbf{from 
$\gcone$ to $\gconetwo$}
is a function $f:\gcone\to \gconetwo$ such that 
\begin{enumerate}
\item If $\alpha,\beta\in \gcone$ then $f(\alpha+\beta)=f(\alpha)+f(\beta)$;
\item If $\alpha\in \gcone$ and $r>0$ then $f(r\alpha)=rf(\alpha)$. 
\end{enumerate}
\begin{examples}
 
The following sets are examples of convex cones. 
\begin{enumerate}
\item Every vector space is a convex cone in itself. 
\item The   intervals $(0,+\infty)$ 
and
$(-\infty,0)$
are convex cones in $\RR$. 

\item The set
$\{x\in\RR^2:x_1>0,x_2>0\}$ is 
a convex cone in $\RR^2$.
 
\item If $V$ is a given vector space and $\alpha\in V\setminus\{0\}$,
the set 
\begin{equation}
\rgb{\alpha}\eqdef\{r\alpha:r\geq0\},
\label{eq:ccgbs} 
\end{equation}
is a convex cone, 
called the  \textbf{ray generated by} $\alpha$. The set
\begin{equation}
\ccgb{\alpha}\eqdef\{r\alpha:r>0\}
\label{eq:ccgbopen} 
\end{equation}
is also a convex cone. 

\item If $\alpha,\beta\in V$ are linearly independent, the set 
$$
\ccgb{\alpha,\beta}=\{r\alpha+s\beta:r,s>0\},
$$
is a convex cone, 
called the  \textbf{angle generated by } $\alpha$ and 
$\beta$ (in $V$). 

\end{enumerate}
\end{examples}

\begin{lemma}
If ${\{\gcone_{i}\}}_{i\in I}$ is a nonempty family of convex cones in a given vector space, and  $\bigcap_{i\in I}\gcone_{i}$ is nonempty, then it is a convex cone. 
\label{l:inters}
\end{lemma}
\begin{proof}
If $\alpha,\beta\in\bigcap_{i\in I}\gcone_{i}$ 
and $r>0$, 
then, for each $i\in I$, $\alpha,\beta\in\gcone_i$, 
hence $\alpha+\beta\in\gcone_i$, thus 
$\alpha+\beta\in\bigcap_{i\in I}\gcone_{i}$. Similarly, 
$r\alpha\in\gcone_i$ for each $i\in I$, hence 
$r\alpha\in\bigcap_{i\in I}\gcone_{i}$.
\end{proof}

Observe that if $V$ is a real vector space and 
$S$ is a nonempty subset of $V$, then the family of 
convex cones in $V$ which contain $S$ is nonempty 
(since $V$ belongs to this family)
and its intersection is nonempty (since it contains $S$). 

\begin{definition}
If $V$ is a real vector space and 
$S$ is a nonempty subset of $V$, then the intersection of 
all convex cones in $V$ which contain $S$ is called the 
\textbf{convex-conical hull} of $S$ and is denoted by $\ccgb{S}$.
\end{definition}

\begin{lemma}
If $V$ is a real vector space and 
$S$ is a nonempty subset of $V$, then 
$\ccgb{S}$ is  the smallest convex cone which contains $S$ and 
\begin{equation}
\ccgb{S}=
\left\{
\alpha\in V:
\,\,
\alpha=\sum_{i=1}^{n}r_i\beta_i,
\,\,
n\in\NN\setminus\{0\},
\,\,
\beta_i\in S,
\,\,
r_i>0
\right\},
\label{eq:ccgbone}
\end{equation}
The sums which appear in~\eqref{eq:ccgbone} are called 
\textbf{conical combinations} of elements of $S$. 
If $S=\{\alpha\}$ then $\ccgb{S}=\ccgb{\alpha}$
as defined in~\eqref{eq:ccgbopen}.
\end{lemma}
\begin{proof}
If $\gcone$ is a convex cone in $V$ which contains $S$ then 
$\gcone$ is a member of the family which is used to define $\ccgb{S}$ and 
hence $\ccgb{S}\subset \gcone$. Now observe that 
the set which appears in the right-hand side of~\eqref{eq:ccgbone} is a convex cone that contains $S$ and which is contained in 
every convex cone which contains $S$. Hence it is equal to 
$\ccgb{S}$. 
\end{proof}

\begin{lemma}
\label{l:inverseimage}
If $f:V\to V$ is a linear map between the 
real vector spaces $V$ and $V'$
and $\gconetwo\subset V'$ is a convex cone in $V'$
then 
the inverse image of $\gconetwo$ along $f$, i.e., the set
$$
f^{-1}(\gconetwo)\eqdef\{
\alpha\in V:
f(\alpha)\in\gconetwo
\}
$$
is a convex cone in $V$.
\end{lemma}
\begin{proof}
If $\alpha,\beta\in f^{-1}(\gconetwo)$ 
and $r>0$, then 
$f(\alpha)$ and $f(\beta)$  belong to $\gconetwo$, 
hence 
$f(\alpha)+f(\beta)\in\gconetwo$, and since 
$f$ is linear, 
$f(\alpha)+f(\beta)=f(\alpha+\beta)$, thus  $f(\alpha+\beta)\in\gconetwo$, i.e., 
$\alpha+\beta\in f^{-1}(\gconetwo)$.
Similarly, $f(r\alpha)=rf(\alpha)\in\gconetwo$, 
hence $r\alpha\in f^{-1}(\gconetwo)$.
\end{proof}
\begin{corollary}
If $k\geq0$ is an integer then 
$\{\alpha\in\polinomia:\alpha(k)\geq0\}$
and
$\{\alpha\in\polinomia:\alpha(k)\leq0\}$ 
are convex cones in $\polinomia$.
\label{c:convexcones}
\end{corollary}
\begin{proof}
Observe that the map $f:\polinomia\to\RR$ defined by 
$f(\alpha)\eqdef\alpha(k)$ is a linear map,
$\{\alpha\in\polinomia:\alpha(k)\geq0\}=f^{-1}([0,+\infty))$,
$\{\alpha\in\polinomia:\alpha(k)\leq0\}=f^{-1}((-\infty,0])$, and 
both $[0,+\infty)$ and $(-\infty,0]$ are convex cones in $\RR$. 
Now apply Lemma~\ref{l:inverseimage}.
\end{proof}

\begin{definition}
If $A$ is a nonempty subset of a given vector space $V$, we denote by 
\begin{equation}
\linearhull{A}\eqdef
\left\{\sum_{i=1}^{n}r_i\beta_i:n\in\NN\setminus\{0\},
r_i\in\RR,\beta_i\in A
\right\}
\label{eq:lhe} 
\end{equation}
the \textbf{linear hull} of $A$. 
If $A=\{\alpha\}$ we write $\linearhull{\alpha}$ for 
$\linearhull{\{\alpha\}}$. 
\end{definition}

Observe that $\linearhull{A}$ is the smallest vector subspace of 
$V$ which contains $A$. 

\begin{lemma}
If $\gcone$ is a convex cone then 
$$
\linearhull{\gcone}=\gcone+(-\gcone)=
\left\{\sum_{i=1}^{n}r_i\beta_i:n\in\NN\setminus\{0\},
r_i\in\{1,-1\},
\beta_i\in \gcone
\right\}
$$
\label{l:linearhull} 
\end{lemma}

\begin{proof}
It suffices to observe that we may rewrite 
the sum which appear in~\eqref{eq:lhe} by placing together 
the terms with positive coefficients and then, in a separate group, those 
with a  negative one, and that, since $\gcone$ is a convex cone, nothing is lost 
is we only multiply by $1$ or $-1$. 
\end{proof}

\begin{definition}
The \textbf{(linear) dimension} of a convex cone 
$\gcone$
is the dimension 
of the linear hull of $\gcone$.
\end{definition}
For example, an open ray has dimension one, 
and an open angle has dimension two. 

\begin{definition}
If $\gcone$ is a convex cone, the set 
$$
\lineal{\gcone}=\{0\}\cup(\gcone\cap(-\gcone))
$$ 
is called the  
\textbf{lineality space} of $\gcone$. 
A convex cone $\gcone$
is called 
\textbf{pointed} if $\lineal{\gcone}=\{0\}$. 
\end{definition}
If 
$\gcone$ is a convex cone then $\lineal{\gcone}$ is a vector 
subspace of the ambient space. 
Cf. Schneider (2022). 
Observe that 
if $\gcone$ is a convex cone in a vector space $V$ then the following conditions are equivalent:
\begin{enumerate}
\item $\gcone$ is a pointed convex cone of dimension one.
\item $\gcone$ is a ray in $V$, i.e., there exists 
$\alpha\in V\setminus\{0\}$ and 
$\gcone=\{r\alpha:r\geq0\}$.
\end{enumerate}

\subsection{Extremal Rays}

We now turn our attention to the notion of extremal ray in a given cone. 
\begin{definition}
Let $\gcone$ be a convex cone. 
An \textbf{extremal ray} in  $\gcone$ is a ray 
$\ray$ in $\gcone$ such that 
\begin{description}
\item[(a)] $\ray\setminus\{0\}\subset\gcone$

\item[(b)] if $\beta,\gamma\in\gcone\setminus\{0\}$
and 
$\beta+\gamma\in\ray\setminus\{0\}$
then $\beta,\gamma$ are linearly dependent.

\end{description}
We denote by $\extremal{\gcone}$ the collection
 of all extremal rays in $\gcone$. 
\end{definition}

Observe that if $\ray$ is an extremal ray of a convex cone $\gcone$, 
$\beta,\gamma\in\gcone\setminus\{0\}$
and $\beta+\gamma\in\ray\setminus\{0\}$, then 
$\{\beta,\gamma\}\cap\ray\not=\emptyset$.

\begin{example}
If $\gcone=\{x\in\RR^3:x_1\geq0,x_2\geq0,x_3\geq0\}$, 
$\ray_1=\{x\in\RR^3:x_1=x_2=x_3\geq0\}$,
and
$\ray_2=\{x\in\RR^3:x_1\geq0,x_2=x_3=0\}$,
then 
$\ray_1\not\in\extremal{\gcone}$
and
$\ray_2\in\extremal{\gcone}$. 
\end{example}

\begin{example}
If $\gcone\eqdef\{x\in\RR^2:x_1>0,x_2>0\}$ then 
$\extremal{\gcone}=\emptyset$. 
\label{eg:noextremalray}
\end{example}

\begin{example}
Let $\gcone\eqdef\{x\in\RR^2:x_2\geq0\}$, 
$\ray_1\eqdef\{x\in\RR^2:x_1\leq0,x_2=0\}$,
and
$\ray_2\eqdef\{x\in\RR^2:x_1\geq0,x_2=0\}$.
Then $\extremal{\gcone}=\{\ray_1,\ray_2\}$.
\label{eg:flat}
\end{example}
The following example will enable us to better appreciate 
the notion of \textit{admissible convex cone}, to be 
presented momentarily. 
\begin{example}
Define $S=\{\alpha_k:k=1,2,3,4\}$ as follows:
$\alpha_1\eqdef(1,1,0)$,
$\alpha_2\eqdef(1,-1,0)$,
$\alpha_3\eqdef(1,0,1)$, 
$\alpha_4\eqdef(1,0,-1)$.
Then $S\subset\RR^3$. Let $\gcone\eqdef\ccgb{S}$.
Then $\gcone$ is a pointed cone, 
$\extremal{\gcone}=\{\rgb{\alpha_1},
\rgb{\alpha_2},
\rgb{\alpha_3},
\rgb{\alpha_4}\}$, and 
$\alpha_1+\alpha_2=\alpha_3+\alpha_4=(2,0,0)$.
\label{eg:notli}
\end{example}

\begin{lemma}
In the vector space $\RR[[\tempo]]$ of formal power series in 
the indeterminate $\tempo$, 
define the convex cone $\gcone$ 
by 
$\gcone\eqdef\{\alpha:\RR[[\tempo]]:\alpha(i)\geq0
\text{ for each } i\in\NN\}$, and let, for each $i\in\NN$, denote by 
$\ray_i$ the ray generated by $\tempo^i$, 
i.e., 
$\ray_i\eqdef\rgb{\tempo^i}=\{\alpha\in\RR[[\tempo]]:\alpha(i)\geq0,
\alpha(j)=0\,\, \forall j\not=i \}$.
Then 
$$
\extremal{\gcone}=\{\ray_i:i\in\NN\}.
$$
\label{eg:fps}
\end{lemma}
\begin{proof}
Firstly, we show that for each $i\in\NN$, 
the ray $\ray_i$ is an extremal ray in $\gcone$.
Indeed, assume that $\beta,\gamma\in\gcone\setminus\{0\}$
and $\beta+\gamma\in\ray_i\setminus\{0\}$.
Let $j\not=i$. Then $\beta(j)+\gamma(j)=0$
(since $\beta+\gamma\in\ray_i$), hence 
$\beta(j)=\gamma(j)=0$
(since $\beta(j)\geq0$ and $\gamma(j)\geq0$).
Hence $\beta,\gamma\in\ray_i$.
It follows that $\ray_i$ is extremal.

Now we show that if $\ray$ is an extremal ray in $\gcone$
then there exists $i\in\NN$ such that $\ray=\ray_i$. Since 
$\ray$ is a ray in $\gcone$, there exists 
$\alpha\in\gcone\setminus\{0\}$
such that $\ray=\rgb{\alpha}$.  Since $\alpha\not=0$, 
the set $\{k\in\NN:\alpha(k)>0\}$ is not empty. 
Let $i$ be the minimum of this set. Let us assume that 
there exists $j>i$ such that $\alpha(j)>0$.
Define 
$$\beta\eqdef\alpha-\alpha(j)\tempo^j,
\gamma\eqdef\alpha(j)\tempo^j$$
Then $\beta,\gamma\in\gcone\setminus\{0\}$ and
$\beta+\gamma=\alpha$. Since $\ray$ is extremal, 
$\beta$ and $\gamma$ are linearly dependent, hence 
there exists $s\not=0$ such that $\beta=s\gamma$.
Hence $\alpha(i)=\beta(i)=s\gamma(i)=0$, a contradiction. 
Hence $\ray=\ray_i$.
\end{proof}

\subsection{Admissible Convex Cones}

\begin{definition}
Let $\gcone$ be a convex cone. We say that 
$\gcone$ is 
\textbf{admissible} 
if $\primo$ $\extremal{\gcone}\not=\emptyset$  
and $\secondo$
each $\alpha\in\gcone$ may be written in a unique way in the form 
\begin{equation}
\alpha=\sum_{\ray\in\extremal{\gcone}}
\nproj{\alpha}{\ray}{\gcone} 
\label{eq:newsumnew} 
\end{equation}
where 
\begin{description}
\item[(a)] 
For each $\ray\in\extremal{\gcone}$,
the element
$\nproj{\alpha}{\ray}{\gcone}$, called 
\textbf{the $\ray$-component of $\alpha$ in $\gcone$},
belongs to $\ray$ for  each 
$\alpha\in\gcone$.

\item[(b)]
For each $\alpha\in\gcone$, the set 
\begin{equation}
\supp{\alpha}\eqdef\{\ray\in\extremal{\gcone}:
\nproj{\alpha}{\ray}{\gcone}\not=0\}
\label{eq:support} 
\end{equation}
(called the 
\textbf{support} of $\alpha$)
is finite.
\end{description}

\label{d:rcc}
\end{definition}

The notion of \textit{support} introduced in~\eqref{eq:support} 
should not be confused with the one in~\eqref{eq:tsupport}.

\begin{definition}
The expression~\eqref{eq:newsumnew} is called the 
\textbf{analysis of $\alpha$ in its extremal projections}.  
\label{d:extremalprojections}
\end{definition}

\begin{example}
The cone $\gcone$ defined in 
Example~\ref{eg:noextremalray}
is not admissible, since $\extremal{\gcone}=\emptyset$.
\end{example}

\begin{example}
The cone $\gcone$ defined in Example~\ref{eg:flat}
is not admissible, since 
the sum of one element of 
$\ray_1$ with an element of $\ray_2$ is confined to 
$\{x\in\RR^2:x_2=0\}$.
\end{example}
\begin{example}
The cone $\gcone$ in Example~\ref{eg:notli}
is not admissible, since $(2,0,0)\in\gcone$ may be written in two 
ways as $\alpha_1+\alpha_2$ or as $\alpha_3+\alpha_4$. 
\end{example}

\begin{example}
The cone $\gcone$ considered in Lemma~\ref{eg:fps}
is not admissible, since the element  $\alpha\in\gcone$
defined by 
 $\alpha(j)=1$ for each 
$j\in\NN$ cannot be written as a finite sum of 
elements that belong to the extremal rays. 
\end{example}

\begin{example}
If $\gcone\eqdef\{x\in\RR^2:x_1\geq0,x_2\geq0\}$
then $\gcone$ is an admissible convex cone. Indeed, its extremal rays are the rays $\ray_1\eqdef\{x\in\RR^2:x_1\geq0,x_2=0\}$,
$\ray_2\eqdef\{x\in\RR^2:x_1=0,x_2\geq0\}$ and the projections are defined by $\nproj{x}{\ray_1}{\gcone}=x_1$,
$\nproj{x}{\ray_1}{\gcone}=x_1$.  
\end{example}

The proof of the following result is left to the reader. 
\begin{lemma}
If   $\gcone$ is an admissible convex cone, then 
$\extremal{\gcone}\not=\emptyset$ 
and there exists a unique map 
(called \textbf{projection} of $\gcone$ along the extremal rays of 
$\gcone$)
\begin{equation}
\begin{split}
\mbox{}
\, & \extremal{\gcone}\times\gcone\to 
\gcone \\
&\,\,\,\,\,\,\, (\ray,\alpha)\mapsto \,\nproj{\alpha}{\ray}{\gcone}
\end{split}
\label{eq:projection}
\end{equation}
such that~\eqref{eq:newsumnew} holds, where only 
finitely many elements in the sum are different from zero, and 
 (i) for each $\ray\in\extremal{\gcone}$,
the element
$\nproj{\alpha}{\ray}{\gcone}$
belongs to $\ray$ for  each 
$\alpha\in\gcone$, 
 (ii) 
the map 
$\gcone\to\ray$, 
defined by 
$\alpha\mapsto \nproj{\alpha}{\ray}{\gcone}$,
is a 
$\ccones$-homomorphism.
\end{lemma}

An example of an admissible convex cone, which plays an important role 
in the subject, will be presented momentarily.
The interest of the notion of \textit{admissible convex cone} 
is due to the existence of a 
\textit{canonical} or 
\textit{intrinsic} decomposition,  given  in~\eqref{eq:newsumnew}. As we will see momentarily, 
in the applications of this notion 
to the subject matter of the present contribution,
the intrinsic decomposition described in~\eqref{eq:newsumnew}
may be seen as the mathematical counterpart of the Italian legislation 
related to lending.

\section{Transactions, Loans, and Arbitrages}

The notion of a \textit{transaction} 
is introduced 
in our treatment for technical reasons: It is loose enough to include 
a wide variety of different cases, not all of which 
correspond 
to what we would call a \textit{loan}. 

\begin{definition}
A \textbf{transaction} is a financial operation 
$\alpha\in\polinomia$ which has the property that 
$$
\alpha(0)\leq0
\,\text{ and }
\alpha(k)\geq0
\,\text{ for all }
k\geq 1.
$$ 
\end{definition}
\begin{lemma}
The collection 
$$
\ploans\eqdef\{\alpha\in\polinomia: 
\alpha
\text{ is a transaction}\}
$$ is a convex cone in $\polinomia$. 
\label{l:ploansisaconvexcone}
\end{lemma}
\begin{proof}
Consider $A_0\eqdef\{\alpha\in\ploans:\alpha(0)\leq0\}$ 
 and, for $k\geq1$,
 $A_k\eqdef\{\alpha\in\ploans:\alpha(0)\geq0\}\subset\polinomia$. 
It suffices to apply Corollary~\ref{c:convexcones}
and Lemma~\ref{l:inters}, since 
$\ploans=\bigcap_{k\geq0}A_k$.
\end{proof}
Observe that 
not all of the following examples 
fit within the usual  notion of a \textit{loan}, 
although they all fit within the notion of 
\textit{transaction}.
$$
-100,
\,\,\,\,\,
-100+50\annob, 
\,\,\,\,\,
-100+50\annob+50\annob^2, 
\,\,\,\,\,
-100+100\annob^{10}
\,\,\,\,\,
-100+11\annob+108\annob^2, 
\,\,\,\,\,
20\annob^2
$$
The class of all \textit{loans} will be defined 
momentarily as a proper subclass of the collection of all transactions.
Nevertheless, the terminology that is usually applied in the case of a \textit{loan} will also be used in the case of a transaction. 
Hence if $\alpha\in\ploans$ then  $\vert\alpha(0)\vert=-\alpha(0)$ is called the \textbf{principal} (the amount of money that is lent) to be payed at time $0$, when the contract is signed. If 
$k\geq 1$ then $\vert\alpha(k)\vert=\alpha(k)$ 
denotes the \textbf{installment} which is due precisely after  $k$ years.

If $\alpha$ is a transaction, then $-\alpha$
encodes the cash flow of  a financial operation where one \textit{receives} the principal $\vert\alpha(0)\vert=-\alpha(0)$ 
and \textit{pays} the installments $\vert\alpha(k)\vert$ in due time. Our notion of  a transaction includes the case where $\alpha(k)\equiv0$,
where nothing is payed at all at any time, which corresponds to the zero polynomial, denoted by $0$. 

\subsection{The Convex Cone of Transactions is 
Admissible}

\begin{definition}
\label{eg:deltas}
If $i\geq0$ we define $\delta_i\in\polinomia$ as follows: 
$$
\delta_0\eqdef-1,
\,\,\,\,
\delta_1\eqdef\annob,
\,\,\,\,
\delta_2\eqdef\annob^2,
\,\,\,\,
\delta_3\eqdef\annob^3,
\ldots
$$
Then 
$\delta_i\in\ploans$ for each $i\geq0$.
The ray generated by $\delta_i$ is denoted by 
$\Delta_i$. Hence $\Delta_i\eqdef\rgb{\delta_i}\subset\ploans$. 
\end{definition}

The role of the rays $\Delta_i$ is described in the following result.
\begin{lemma}
$\extremal{\ploans}=\{\Delta_i:i\geq0\}$.
\label{l:extremalt}
\end{lemma}
\begin{proof}
The fact that for each $i\in\NN$ the set 
$\Delta_i$ is a ray in $\ploans$ follows at once from the definition. 
We now show that $\Delta_0$ is an extremal ray in 
$\ploans$. 
Assume that $\beta,\gamma\in\ploans\setminus\{0\}$
and $\beta+\gamma\in\Delta_0\setminus\{0\}$.
Then $\beta(0)+\gamma(0)<0$, 
$\beta(0)\leq0$, $\gamma(0)\leq0$,
and, for each $i\geq1$,
$\beta(i)\geq0$, 
$\gamma(i)\geq0$, and 
$\beta(i)+\gamma(i)=0$. It follows that 
$\beta(i)=\gamma(i)=0$ for each $i\geq1$, 
$\beta(0)<0$, $\gamma(0)<0$, hence $\beta$ and $\gamma$
are linearly dependent. 
We now show that $\Delta_1$ is an extremal ray in 
$\ploans$, the proof for the other rays being similar.
Assume that $\beta,\gamma\in\ploans\setminus\{0\}$,
and
$\beta+\gamma\in\Delta_1\setminus\{0\}$.
Then $\beta(0)+\gamma(0)=0$,
$\beta(1)+\gamma(1)>0$,
and, for each $i\geq2$ 
$\beta(i)+\gamma(i)=0$. It follows that 
$\beta(0)=\gamma(0)=0$, and 
$\beta(i)=\gamma(i)=0$ for each $i\geq2$.
It follows that $\beta$ and $\gamma$ are linearly dependent. 

We now show that if $\ray$ is an extremal ray in $\ploans$
then there exists $i\in\NN$ such that 
$\ray=\Delta_i$. Since $\ray$ is a ray, there exists $\alpha\in\ploans\setminus\{0\}$ such that 
$\{r\alpha:r>0\}=\ray$.
Since $\alpha\not=0$, there exists $k\in\NN$ such that 
$\alpha(k)\not=0$. Let us assume that there exists 
$j\in\NN\setminus\{k\}$ such that $\alpha(j)\not=0$.
We may assume, without loss of generality, 
that $0\leq k<j$, and then define
$$
\beta\eqdef\alpha-\alpha(j)\delta_j,\gamma\eqdef\alpha(j)\delta_j
$$
Then 
$$
\beta,\gamma\in\ploans\setminus\{0\},
\beta+\gamma=\alpha\in\ray
$$
Since $\ray$ is extremal, it follows that $\beta$ and $\gamma$
are linearly dependent, hence there exists $s\not=0$ such that 
$\beta=s\gamma$. In particular, $\alpha(k)=\beta(k)=s\gamma(k)=0$, 
a contradiction. It follows that for each $j\not=k$ $\alpha(j)=0$.
Hence $\ray=\Delta_k$.
\end{proof}
\begin{proposition}
The convex cone $\ploans$ is admissible. 
\end{proposition}
\begin{proof} In Lemma~\ref{l:ploansisaconvexcone}
and in Lemma~\ref{l:extremalt}
we have already shown that $\ploans$
is a convex cones and that $\extremal{\ploans}\not=\emptyset$, since 
$\extremal{\ploans}=\{\Delta_i:i\in\NN\}$.  
Hence 
the property described in $\primo$ in 
Definition~\ref{d:rcc} holds.
Now consider the map 
\begin{equation}
\begin{split}
\mbox{}\, & \extremal{\ploans}\times\ploans\to 
\ploans \\
&\,\,\,\,\,\,\, (\alpha,\Delta_i)\mapsto \,
\nproj{\alpha}{\Delta_i}{\ploans} 
\end{split}
\end{equation}
defined by 
$$
\nproj{\alpha}{\Delta_i}{\ploans} \eqdef\alpha(i)\annob^i
$$
The fact that each $\alpha\in\ploans$ 
is a polynomial implies at once that 
the support of each $\alpha\in\ploans$
is finite and that the property described in $\secondo$ in 
Definition~\ref{d:rcc} holds. 
\end{proof}

\subsection{Loans}

\begin{definition}
A transaction $\alpha\in\ploans$ is called a \textbf{loan} if
\begin{equation}
\sum_{k\geq1}\alpha(k)\geq -\alpha(0)>0
\label{eq:properloans} 
\end{equation}
 \end{definition}
Observe that the sum in~\eqref{eq:properloans} is well-defined, since only finitely many terms are nonzero.
The meaning of~\eqref{eq:properloans} 
is that the sum of the installments is 
greater than or equal to  the principal, and the principal is nonzero.
For example, 
according to this definition, 
$-100+50\annob$ is \textit{not} a loan, while
$-100+100\annob$ is a loan.

\paragraph{Interpretation.}
In our model, to \textit{buy} a loan means to \textit{lend} the principal and \textit{receive} the installments;
to \textit{sell} a loan means to \textit{borrow} the principal and \textit{pay} the installments.

\begin{lemma}
The collection 
$$
\loans\eqdef\{\alpha:\alpha\in\polinomia,
\,
\alpha
\text{ is a loan}
\}
$$
is a convex cone in $\polinomia$. 
\end{lemma}

\begin{proof}
Left to the reader.  
\end{proof}

Observe that 
$$
\loans\subset\ploans\subset\polinomia
$$

\subsubsection{Elementary Loans}

\begin{definition}
Let $n\in\NN\setminus\{0\}$.
If 
$\alpha\in\ploans$ may be written in the form 
\begin{equation}
\alpha=\alpha(0)+\alpha(n)\annob^n 
\label{eq:elomn}
\end{equation}
where $\alpha(0)$ and $\alpha(n)$ are both nonzero, then 
$\alpha$ is called an 
\textbf{elementary loan} (of maturity $n$).
 We denote by $\element{n}$ 
the collection of elementary loans of maturity $n$.  
\end{definition}
Hence 
$$
\element{n}=\{-c+d\annob^n:c,d>0\}
$$
\begin{lemma}
The collection of elementary loans of maturity $n$ 
is a convex cone in $\polinomia$.
\end{lemma}

\begin{proof}
Left to the reader.
\end{proof}

\subsubsection{The Net (Nominal) Gain Operator}

The meaning of~\eqref{eq:properloans} is that the sum 
of the installments is at least equal to the principal. 
Indeed, if $\alpha$ is a loan, then 
the sum 
\begin{equation}
\sum_{k\geq0}\alpha(k)
\label{eq:sum}
\end{equation}
equals the difference 
between the sum of the installments and the principal, i.e., it is the \textbf{net (nominal) gain} obtain by the lender through the loan 
$\alpha$. For example, if $\alpha$ is 
the loan in~\eqref{eq:firstexample} or the loan 
in~\eqref{eq:secondexample}, then 
the net gain is equal to $20$ in both cases. 
The sum in~\eqref{eq:sum} is thus equal to the quantity 
obtained by the lender \textit{beyond the principal}, which  in Roman Law was denoted as 
\textit{quid ultra sortem}.
The net gain 
associated to $\alpha$, expressed by~\eqref{eq:sum}, is denoted by 
$\net(\alpha)$. Hence $\net$ is the operator 
$$
\net:\loans\to\RR,
\,
\text{ defined by }
\,
\net(\alpha)\eqdef\sum_{k\geq0}\alpha(k),
$$
In our model of a loan market, as we will see, 
$\net(\alpha)$ is a rather rough measure of the \textit{effective} gain
obtained by the lender. For example, the loan in~\eqref{eq:firstexample} is actually \textit{more convenient} to the lender than 
the loan 
in~\eqref{eq:secondexample}, 
although they have the same nominal gain, 
since in~\eqref{eq:firstexample} the lender may lend $10$ 
at time 1
for one year and finally obtain (at time $2$) an amount which is \textit{greater} that 
$120$ (the precise amount will depend on the interest rate available at time 1). Indeed, 
in looking at the mere sum $\sum_{k\geq0}\alpha(k)$ we pretend that the various installments live at time zero, 
which is contrary to the basic tenet of this subject, where, 
as we said, 
money does not live 
\textit{outside} of the temporal dimension ---
an understanding already known in the past under the saying 
\textit{plus dat qui cito dat} (Latin for
\textit{the sooner you give, the more you give}). 
We shall return on this point after we introduce our model of a loan market.
For the time being, we emphasize the fact that 
the notion of \textit{net gain} is a very rough measure of the gain 
obtained by the lender, since, as we stated at the beginning, 
a basic tenet in the subject is that each amount of money 
is \textit{embedded 
in a time position} (to wit: The time when it is due), while the net gain, as here defined,  
is a mere amount, not imbedded in any particular time. 

We will see that the proper way to evaluate the gain involved 
in a \textit{rewarding loan} is to write the loan, in a canonical way, as the sum of a mutuum and an \textit{arbitrage}. Firstly, we need to clarify these notions.

\subsection{The Notion of Arbitrage}

Our notion of transaction also includes the somewhat unnatural case where $\alpha$ is an \textit{arbitrage}.

\begin{definition}
An \textbf{arbitrage} is a transaction 
$\alpha\in\polinomia$
such that
$\alpha(0)=0$ and 
$\net(\alpha)>0$.
\end{definition}

\begin{example}
The transaction $10\annob$ 
is an arbitrage. 
\end{example}

\paragraph{Interpretation.}
In an arbitrage, one receives a positive income in installments, without going through the trouble of lending any money. In our definition, an arbitrage is a transaction but not a loan. 

\begin{lemma}
The collection 
$$
\arbit\eqdef\{\alpha:\alpha\in\loans,
\alpha
\text{ is an arbitrage}
\}
$$
is a convex cone in $\polinomia$. 
\end{lemma}

\begin{proof}
Left to the reader. 
\end{proof}

\subsubsection{Rewarding Loans}

\begin{definition}
A transaction $\alpha$ is called a \textbf{rewarding loan} if 
$$
\sum_{k\geq1}\alpha(k)>-\alpha(0)>0
$$
\end{definition}
Hence a rewarding loan is a loan, and 
$-100+100\annob$ is a loan which is not rewarding, while 
$-100+101\annob$ is a rewarding loan.

\begin{lemma}
The collection 
$$
\rloans\eqdef\{\alpha\in\polinomi:
\alpha
\text{ is a rewarding loan }\}
$$ 
is a convex cone in 
$\polinomia$. 
\end{lemma}

\begin{proof}
Left to the reader.  
\end{proof}

Observe that 
$$
\rloans\subset\loans
$$

\begin{remark}
An elementary transaction is not necessarily a  loan. For example, 
$$
-100+10\annob^2
$$ 
is an elementary transaction but it is not a loan. 
From the point of view of the mere cash flow, an elementary transaction which is a rewarding loan has the same effect as those investment instruments known as \textit{zero coupon bonds}.
We prefer to adopt a different terminology because of  the difference 
 in Italian legislation 
between 
loans (such as consumer loans) and investment instruments 
(a  distinction which is akin to the one between commercial banking and investment banking). 
\end{remark}

\subsubsection{Mutua}

\begin{definition}
A loan $\alpha$ is called a \textbf{mutuum} if 
\begin{equation}
\sum_{k\geq1}\alpha(k)=-\alpha(0)>0
\label{eq:mutuum}
\end{equation}
 \end{definition}
The proof of the following result is left to the reader. 
\begin{lemma}
The collection 
$$
\mutuum
\eqdef
\{\alpha\in\polinomi:
\alpha
\text{ is a mutuum}\}
$$ 
is a convex cone in $\polinomi$.
\end{lemma}
Observe that 
$$
\mutuum\subset\loans
\text{ and }
\mutuum\cap\rloans=\emptyset
$$
The meaning of~\eqref{eq:mutuum} 
is that the sum of the installments is equal to the principal, 
as in a \textit{mutuum} in Roman Law, where 
the eventual agreement on the payment of  interests was made under a different legal instrument
(the \textit{stipulatio usurarum}). 
For example, the following loans are all  mutua:
\begin{equation}
-100+50\annob+50\annob^2,
\,\,\,\,\,
-100+30\annob+70\annob^2,
\,\,\,\,\,
-100+100\annob,
\,\,\,\,\,
-100+100\annob^2,
\,\,\,\,\,
-100+100\annob^3
\label{eq:examplesmutua} 
\end{equation}
In the following diagram all inclusions 
are proper (and the empty set appears as the intersection of 
$\mutuum$ and $\rloans$).
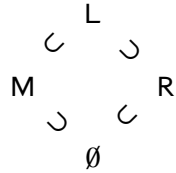
\[
\rotatebox{45}{$
\begin{array}{ccc}
\rotatebox{-45}{$\mutuum$} & \subset & \rotatebox{-45}{$\loans$} \\
\rotatebox{-90}{$\supset$}& &\rotatebox{-90}{$\supset$}\\[9pt]
\rotatebox{-45}{$\emptyset$} & \subset &\rotatebox{-45}{$\rloans$}
\end{array}
$}
\]

In the last three examples in~\eqref{eq:examplesmutua}, the principal is payed back 
in just one installment. This kind of mutuum deserves a special name.
\begin{definition}
If $n\in\NN\setminus\{0\}$ then 
a mutuum of the form 
$$
-c+c\annob^n
$$ 
for some $c>0$ is called an \textbf{elementary mutuum} of maturity $n$ and principal $c$. 
\end{definition}

\begin{lemma}
An  elementary rewarding loan  may be uniquely written as the sum 
of an elementary mutuum and an arbitrage. 
\label{l:uniquely}
\end{lemma}
\begin{proof}
It suffices to observe that $-c+d\annob^n$, where $c>0$
and $d>c$, may be written as the sum of 
$$ 
(-c+c\annob^n)+[(d-c)\annob^n]
$$
where $-c+c\annob^n$ is a mutuum and 
$(d-c)\annob^n\in\arbit$, and this expression is unique.
\end{proof}

\begin{example}
For example, the elementary rewarding loan $-50+60\annob^2$ may be uniquely written 
as the sum of a mutuum and an arbitrage, as follows.
$$
-50+60\annob^2=(-50+50\annob^2)+10\annob^2
$$ 
\end{example}

\paragraph{A general problem in the background.}
The general problem in the background is that of writing a rewarding loan as the sum 
of a mutuum and an arbitrage in some \textit{intrinsic} way. Whenever this can be done, 
it is possible to consider the arbitrage component of a given rewarding loan as the 
“interest” involved in the given loan. In other words, the proper notion of “interest” 
should not be seen as a mere amount (the “net gain”), but as an arbitrage which is 
the counterpart that corresponds to the mutuum, as in the following symbolic 
expression
\begin{equation}
\text{ rewarding loan }
=
\text{ mutuum }
+
\text{ arbitrage } 
\label{eq:symbolic}
\end{equation}
In other words, if 
a rewarding loan can be  written 
in an intrinsic way as the sum of a mutuum and an arbitrage, then the arbitrage 
is the “interest” (not to be confused with the net gain). 
Lemma~\ref{l:uniquely} shows that this task has a clear positive solution if 
the rewarding loan is elementary. This fact shows the importance of being able 
to write a loan (belonging to a certain class) in a unique way as the sum of 
elementary rewarding loans, for which~\eqref{eq:symbolic} can be done 
in only one way. We will return to this point momentarily.

\section{Loan Markets}\label{s:loanmarkets}

\paragraph{Motivation.}
Our notion of \textit{loan market}   
encodes the collection of all loans that are available in the market 
to the economic agents 
at a given time. 
Recall that 
the sum of two polynomials is a representation of the total cash flow that results from the superposition of the corresponding financial operations, and multiplication by a scalar is a representation of the total cash flow that results from acting with many copies of a given financial operation.
These facts show that the notion of convex cone arises in a natural way in the subject.
An extensive introduction to General Equilibrium Theory can be found in 
Ingrao and Israel (1996).

\begin{definition}
A \textbf{loan market} is a convex cone in $\polinomia$ which is contained in 
$\loans$.
\end{definition}

\begin{examples}
If $\alpha\in\loans$ then 
 $\ccgb{\alpha}=\{r\alpha:r>0\}$ is a loan market.
If $\alpha,\beta\in\loans$ are linearly independent, then 
$\ccgb{\alpha,\beta}=\{r\alpha+s\beta:r,s>0\}$
is a loan market.
\end{examples}

\paragraph{Assumptions of the Model.}
We assume the familiar (albeit somewhat idealized) \textit{symmetry hypothesis}, i.e.,  
that if a given loan market $\mercato$ contains a loan $\alpha$ then all 
the agents acting in the market have the opportunity to buy or to sell  $\alpha$ (to \textit{buy} a loan means to \textit{lend} the principal and \textit{receive} the installments;
to \textit{sell} a loan means to \textit{borrow} the principal and \textit{pay} the installments). 
Moreover, we assume that the loan market will change over time, 
more precisely, we assume that the agents will make their investment choices periodically, 
according to 
 the time unit used in our polynomial notation, and that 
 the loan market available to them will change accordingly (in a non predictable way) in that time frame.  
We also assume that there 
is no insolvency risk and that there 
are no transaction costs.
On the basis of this interpretation, the following notion is essential.

\begin{definition}
The \textbf{output} of a loan market 
$\mercato$ is the collection of all financial operations 
that result 
from buying or selling a finite number of loans available in 
$\mercato$. 
\end{definition}
\begin{lemma}
The output of \/ $\mercato$ is 
encoded by 
$\linearhull{\mercato}$, 
the vector space generated in $\mercato$.
\end{lemma}
\begin{proof}
It suffices to observe that the sale of a loan corresponds to change its sign, 
and apply Lemma~\ref{l:linearhull}.
\end{proof}

The following example shows the kind of 
outcome we should like to avoid.
\begin{example}
Consider the following loans:
$$
\alpha=-1+5\annob,\,\,\beta=-1+2\annob
$$
and let $\mercato\eqdef\ccgb{\alpha,\beta}$ be the loan market generated by 
$\alpha$ and $\beta$.  
Then $\linearhull{\mercato}$ contains $3\annob$, which is an arbitrage.  
\label{eg:arbitrage}
\end{example}
The presence of an arbitrage in the output of the loan market $\mercato$ of Example~\ref{eg:arbitrage} 
means that $\mercato$ \textit{assigns prices in an inconsistent way}.

\begin{definition}
A transaction market $\mercato$ is \textbf{arbitrage-free} if 
its output $\linearhull{\mercato}$
contains no arbitrage.  
\end{definition}
\begin{example}
If $\alpha\in\loans$ then 
$\{r\alpha:r>0\}$ is  arbitrage-free.
\end{example}

\paragraph{A Natural Question.}

A natural question is whether the loan market 
generated by a given collection of loans 
is arbitrage-free. 

\begin{definition}
If $Q$ is a collection of loans, i.e., $Q\subset\loans$, we say that 
$Q$ is arbitrage-free if the loan market  
$\ccgb{Q}$ generated by $Q$ is arbitrage-free. 
\label{d:arbitrage-free}
\end{definition}


\subsection{Elementary Loans at Simple Interest}

The following notion is well-established in literature and 
it has been known for many centuries (albeit under different names). 
For example, it is mentioned in the \textit{Summa} by Luca Pacioli, first published in 1494, but it had actually been known and applied at least since the 13th century.

\begin{definition}
An elementary loan $\alpha$ of maturity 
 $n\in\NN\setminus\{0\}$ has (yearly) \textbf{simple interest rate} equal to $x>0$ if 
$$
\alpha(n)+\alpha(0)\cdot(1+nx)=0
$$
\end{definition}
If $x>0$ and $n\in\NN\setminus\{0\}$, 
we denote by $\elemd{x}{n}$ 
the collection of elementary loans of maturity $n$
which have simple interest rate equal to $x$. 
Observe that if $\alpha\in\elemd{x}{n}$ 
then $\alpha$ is an elementary rewarding loan.
Observe that $\elemd{x}{n}\subset\rloans$. 
\begin{lemma}
If $x>0$ and $n\in\NN\setminus\{0\}$, then 
the collection $\elemd{x}{n}$ is a one-dimensional arbitrage-free loan market.
\label{l:elofir}
\end{lemma}

\begin{proof}
Let 
\begin{equation} 
\basis{n}\eqdef-1+(1+nx)\annob^n
\label{eq:elasitwo}
\end{equation}
and 
observe that 
\begin{equation}
\elemd{x}{n}
=
\{c\basis{n}:
c>0
\}
\label{eq:elasi}
\end{equation}
It follows that 
$$
\linearhull{\elemd{x}{n}}
=
\left\{
-c+c(1+nx)\annob^n:c\in\RR
\right\}
$$
and hence $\alpha(0)=0$ implies that $\alpha(n)=0$
for each $\alpha\in\linearhull{\elemd{x}{n}}$.
\end{proof}

\begin{lemma}
If $\alpha\in\elemd{x}{n}$ is written in the form
$-c+c(1+nx)\annob^n$, where $c>0$,
then 
\begin{equation}
\net(\alpha)=cnx
\label{eq:simpleinterest} 
\end{equation}
\end{lemma}

\begin{proof}
$\net(\alpha)=\alpha(0)+\alpha(n)=-c+c(1+nx)=-c+c+cnx=cnx$.
\end{proof}

\paragraph{Interpretation.}
Observe that~\eqref{eq:simpleinterest} describes the
familiar formula for the net gain 
in an elementary loan made at the simple interest rate $x>0$. Indeed, 
since 
$$
c\basis{n}=-c+(c+cnx)\annob^n
$$
the coefficient $c$ plays the role of the principal, while $cnx$ is the net gain. Another way 
to put it is this:
$$
-c+(c+cnx)\annob^n=(-c+c\annob^2)+(cnx\annob^n)
$$
and this is a particular case of Lemma~\ref{l:uniquely}.

\begin{lemma}
If $x>0$ then 
the collection
$$
\elem{x}
\eqdef
\{
\alpha\in\loans:
\alpha
\text{
is an elementary loan that
has simple interest rate equal to 
}
x
\}
$$ 
is a cone but it is not convex.
\label{l:elemloans}
\end{lemma}
\begin{proof}
Observe that 
\begin{equation}
\elem{x}=\bigcup_{\ell=1}^{\infty}\elemd{x}{\ell}.
\label{eq:union} 
\end{equation}
Hence Lemma~\ref{l:elofir} implies at once that 
$\elem{x}$ is a cone. 
Now consider
$$
\alpha\eqdef-1+(1+x)\annob
\quad
\text{ and }
\beta\eqdef-1+(1+2x)\annob^2
$$
Then $\alpha,\beta\in\elem{x}$ but
$\alpha+\beta=-2+
(1+x)\annob+
(1+2x)\annob^2\not\in\elem{x}$.
\end{proof}

In order to specialize the natural question which we asked 
at the beginning of Section~\ref{s:loanmarkets} 
to the case of the collection of 
elementary loans of a given simple interest rate, we need to 
obtain a sufficient understanding of the convex cone 
generated by these elementary loans.
We will do it in the following section, 
where we will show that this convex cone is admissible. 
We will then proceed to 
show that this convex cone is arbitrage-free 
in the sense of Definition~\ref{d:arbitrage-free}.

\subsection{The Convex Cone Generated by Elementary Loans of 
a Given Simple Interest Rate is Admissible}


We have seen in Lemma~\ref{l:elemloans} that if $x>0$
then  $\elem{x}$ is not a convex cone, hence it is natural to consider the 
convex cone generated by $\elem{x}$. 
\begin{definition}
Let $x>0$.
The convex cone  
generated by $\elem{x}$ is denoted 
by  $\sloans{x}$ and called the \textbf{convex cone of loans of simple 
interest rate $x$}. If $\alpha\in\sloans{x}$ then we say that 
$\alpha$ is a \textbf{loan of simple interest rate} $x$.
\end{definition}
\begin{lemma} If $x>0$ then 
$$
\sloans{x}
\eqdef
\ccgb{\elem{x}}
=
\left\{
\alpha\in\polinomia:
\,
\alpha=\sum_{i=1}^{k}
\beta_i,
\,
k\in\NN\setminus\{0\},
\,
\beta_i\in \elem{x}
\right\}
$$ 
\end{lemma}
\begin{proof}
Since $\elem{x}$ is a cone, nothing is added if we multiply 
by positive numbers. 
\end{proof}
Recall that $\basis{n}=-1+(1+nx)\annob^n$ has been defined in~\eqref{eq:elasitwo}, for 
$n\in\NN\setminus\{0\}$.

The following result says that any conical combination of elementary 
loans of simple interest rate $x$ belongs to $\sloans{x}$.

\begin{lemma}
If $\alpha\in\polinomi$ may be written in the form 
\begin{equation}
c_1\basis{j_1}+c_2\basis{j_2}+c_3\basis{j_3}+\ldots+c_n\basis{j_n} 
\label{eq:sumofbasisZERO}
\end{equation}
where 
\begin{equation}
n\in\NN\setminus\{0\},
1\leq j_1< j_2<\ldots<j_n,
\text{ and }
c_k>0, k=1,2,\ldots,n 
\label{eq:sumofbasisZERObis}
\end{equation}
then $\alpha\in\sloans{x}$. 
\end{lemma}

\begin{proof}
It suffices to observe that the loans
$\basis{j_k}$ belong to $\elem{x}$.  
\end{proof}

The following result says that any 
any element of $\sloans{x}$ may be written as 
conical combination of elementary 
loans of simple interest rate $x$.

\begin{lemma}
If $\alpha\in\sloans{x}$ then 
$\alpha$ may be written in the form~\eqref{eq:sumofbasisZERO}
where~\eqref{eq:sumofbasisZERObis} holds.
\label{l:unique}
\end{lemma}

\begin{proof}
If $\alpha\in\sloans{x}$ then  
$\alpha=\sum_{i=1}^{k}\beta_i$ where
$k\in\NN\setminus\{0\}$, and $\beta_i\in \elem{x}$. 
Observe that~\eqref{eq:union} implies that  
each $\beta_i$ has a certain maturity $m_i$, since the sets which appear in the sum in~\eqref{eq:union} are disjoint. Indeed, $\beta_i$ has maturity $m_i$ precisely if 
$\beta_i\in \elemd{x}{m_i}$. The set 
$\{m_i:i=1,2,\ldots,k\}$ is a finite set $I$ of positive integers, and we may write
$$
\alpha=\sum_{i=1}^{k}\beta_i=\sum_{\ell\in I}\sum_{m_i=\ell}\beta_i
$$
Now each sum $\sum_{m_i=\ell}\beta_i$ is equal to an element of $\elemd{x}{\ell}$, 
and hence, by Lemma~\ref{l:elofir}, it may be written in the form $d_{\ell}\basis{\ell}$
for some $d_{\ell}>0$. Hence 
\begin{equation}
\alpha=\sum_{\ell\in I}d_{\ell}\basis{\ell}
\label{eq:sum1} 
\end{equation}
Now the set $I$ may be written as $I=\{j_1,j_2,\ldots,j_n\}$ with  
$1\leq j_1< j_2<\ldots<j_n$ and $n\in\NN\setminus\{0\}$
in only one way, and~\eqref{eq:sumofbasisZERO} follows. 
\end{proof}

\begin{lemma}
If $\alpha\in\sloans{x}$ may be written in the form~\eqref{eq:sumofbasisZERO}
where $n\in\NN\setminus\{0\}$, 
$1\leq j_1< j_2<\ldots<j_n$, and the constants $c_1, c_2,\ldots, c_n$ 
are strictly positive, then 
\begin{equation}
c_k=\frac{\alpha(j_k)}{1+xj_k}
\label{eq:formulas} 
\end{equation}
\begin{equation}
\sum_{k=1}^{n}c_k=-{\alpha(0)}
\label{eq:formulastwo} 
\end{equation}
and
\begin{equation}
\tsupp{\alpha}=\{0,j_1,j_2,\ldots,j_n\} 
\label{eq:whatisthetsupport}
\end{equation}

\label{l:uniquezero}
\end{lemma}

\begin{proof}
\eqref{eq:formulastwo} follows at once by evaluating $\alpha(0)$. Indeed, 
if $\alpha$ is equal to~\eqref{eq:sumofbasisZERO} then $\alpha(0)$ is equal to 
$$
(c_1\basis{j_1}+c_2\basis{j_2}+c_3\basis{j_3}+\ldots+c_n\basis{j_n})(0)
$$
which is equal to 
$$ 
c_1\basis{j_1}(0)+c_2\basis{j_2}(0)+c_3\basis{j_3}(0)+\ldots+c_n\basis{j_n}(0)
=-\sum_kc_k
$$
In a similar way, $\alpha(j_k)$ is equal to 
$$
(c_1\basis{j_1}+c_2\basis{j_2}+c_3\basis{j_3}+\ldots+c_n\basis{j_n})(j_k)
$$
which is equal to
$$
c_1\basis{j_1}(j_k)+c_2\basis{j_2}(j_k)+c_3\basis{j_3}(j_k)+\ldots+c_n\basis{j_n}(j_k)=
c_k(1+xj_k)
$$
Now~\eqref{eq:whatisthetsupport} follows at once.  
\end{proof}

The following result says that
any element of $\sloans{x}$ may be written 
in a unique way 
as 
conical combination of elementary 
loans of simple interest rate $x$.

\begin{proposition}
If $\alpha\in\sloans{x}$ then 
$\alpha$ may be \textbf{uniquely} written in the form~\eqref{eq:sumofbasisZERO} 
where~\eqref{eq:sumofbasisZERObis} holds.
\label{p:unique}
\end{proposition}

\begin{proof}
The conclusion follows at once from Lemma~\ref{l:unique}
and 
Lemma~\ref{l:uniquezero}. 
Indeed, the set 
$\{j_1,j_2,\ldots,j_n\}$, being equal to 
$\tsupp{\alpha}\!\setminus\!\{0\}$,  
is uniquely determined 
by $\alpha$, 
and the values $c_k$ are uniquely determined by~\eqref{eq:formulas}. 
\end{proof}

We will see that the expression~\eqref{eq:sumofbasisZERO} 
given by Proposition~\ref{p:unique}
is the 
analysis of $\alpha$ in its extremal projections, in the sense of Definition~\ref{d:extremalprojections}.

We are now able to give an intrinsic characterization of those loans 
that belong to $\sloans{x}$.

\begin{corollary}
If $\alpha\in\loans$ and $x>0$ then  the following conditions are equivalent.
\begin{description}

\item[(a)]
$\alpha\in\sloans{x}$

\item[(b)] 
\begin{equation}
-\alpha(0)=\sum_{k\geq1}
\frac{\alpha(k)}{1+xk} 
\label{eq:sumloan}
\end{equation}

\item[(c)] $\alpha$ may be uniquely written in the form~\eqref{eq:sumofbasisZERO}
where~\eqref{eq:sumofbasisZERObis} holds.

\end{description}
\label{c:corollaryintrinsic}
\end{corollary}

\begin{proof}
If $\alpha\in\sloans{x}$ then~\eqref{eq:sumloan} follows at once 
from Lemma~\ref{l:uniquezero}. 
Hence (\textbf{a})
implies (\textbf{b}). 
Now let us assume that $\alpha\in\loans$ and that~\eqref{eq:sumloan} holds.
Then~\eqref{eq:properloans} implies that the 
set $\tsupp{\alpha}\!\setminus\!\{0\}$ is not empty, 
hence we may denote its elements by $\{j_1,j_2,\ldots,j_n\}$,
 with $n\in\NN\setminus\{0\}$,
and
$1\leq j_1< j_2<\ldots<j_n$. 
Now \textit{define} 
$c_k$ by~\eqref{eq:formulas}, i.e., $c_k=\frac{\alpha(j_k)}{1+xj_k}$,
and observe that 
\begin{equation}
\alpha=\sum_{i=1}^n{}c_i\basis{j_i}
\label{eq:equality} 
\end{equation}
Indeed, $\alpha(j_k)=(\sum_{i=1}^n{}c_i\basis{j_i})(j_k)$ follows 
from the very definition of $c_k$, given in~\eqref{eq:formulas}. On the other hand 
 $\alpha(0)=(\sum_{i=1}^n{}c_i\basis{j_i})(0)$ follows from~\eqref{eq:sumloan} and~\eqref{eq:formulas}. Hence~\eqref{eq:equality} follows, since both terms 
 have the same value when evaluated at each $i$ in their common support. 
Thus, (\textbf{b})
implies (\textbf{a}). The equivalence between 
 (\textbf{a})
and (\textbf{c}) has been proved in Proposition~\ref{p:unique}.
\end{proof}

\begin{example}
Let $\alpha$ be the loan in~\eqref{eq:firstexample}. 
Then $\alpha\not\in\sloans{0.1}$. 
Indeed, $\alpha\in\sloans{x}$ for $x=0.1047404277$.
\end{example}

\paragraph{Interpretation.}
Corollary~\ref{c:corollaryintrinsic} gives an intrinsic characterization 
of those loans which belong to $\sloans{x}$. 
The fact that the sum of the coefficients 
$c_k$ which appear in~\eqref{eq:sumofbasisZERO} is equal 
to the principal $|\alpha(0)|$, established in~\eqref{eq:formulastwo}, means that 
they play the role of quotas of the principal 
that are payed back with each installment. 
The following result also serves as an additional clarification of this point. Firstly, we need 
to introduce the notion of \textit{partition}.

\begin{definition}
A \textbf{partition of} $P>0$ is a pair $(n,S)$ where $n\in\NN\!\setminus\!\{0\}$ 
and $S$ is a set of $n$ positive numbers $P_1,P_2,\ldots,P_n$ such that 
$$
P=P_1+P_2+\ldots+P_n
$$
\end{definition}

\begin{lemma}
Given $P>0$, a partition  $(n,S)$ of $P$, $n\in\NN\setminus\{0\}$, 
with $P=\{P_1,P_2,\ldots,P_n\}$, 
and a collection $T=\{j_1,j_2,\ldots,j_n\}$ of $n$ positive numbers,
with 
$1\leq j_1< j_2<\ldots<j_n$,
there exists one and only one loan $\alpha\in\sloans{x}$ such that 
$-\alpha(0)=P$, $\tsupp{\alpha}=\{0\}\cup T$, and 
$$
P_k=\frac{\alpha(j_k)}{1+xj_k}
$$
\label{l:partition} 
\end{lemma}

\begin{proof}
The conditions mean that $\alpha=\sum_{k=1}^n P_k\basis{j_k}$. 
\end{proof}

\begin{lemma}
If $x>0$ then $\extremal{\sloans{x}}=\{\rgb{\basis{n}}:n\in\NN\setminus\{0\}\}$. 
\label{l:extremalray}
\end{lemma}

\begin{proof}
Firstly, we show that for each $n\in\NN\setminus\{0\}$ 
the ray $\rgb{\basis{n}}$ is extremal. Indeed, 
 assume that 
$\beta,\gamma\in\sloans{x}\!\setminus\!\{0\}$
and $\beta+\gamma\in\rgb{\basis{n}}$ with $\beta+\gamma\not=0$.
Hence
\begin{equation}
\beta+\gamma
=
-c+c(1+nx)\annob^n
\label{eq:form} 
\end{equation}
for some $c>0$. Moreover, Lemma~\ref{l:unique} implies that 
$\beta$ and $\gamma$ may be written 
in the form~\eqref{eq:sumofbasisZERO}, and then~\eqref{eq:form} implies that 
they both belong to $\rgb{\basis{n}}$. 

We now show that if $\ray\in\extremal{\sloans{x}}$ then there exists 
$k\in\NN\setminus\{0\}$ such that $\ray=\rgb{\basis{k}}$. Firstly, observe that 
there exists $\alpha\in\sloans{x}$ such that 
$\ray=\rgb{\alpha}$ and $\alpha\not=0$. Now apply Lemma~\ref{l:unique} and write 
$\alpha$ in the form~\eqref{eq:sumofbasisZERO}. Let us assume 
that in~\eqref{eq:sumofbasisZERO} more than one term appears, i.e., let us assume that 
$n>1$. Then we may rewrite $\alpha$ as the sum of 
$\beta\eqdef c_1\basis{j_1}$
and
$\gamma\eqdef c_2\basis{j_2}+c_3\basis{j_3}+\ldots+c_n\basis{j_n}$, and since  
$\ray$ is extremal it follows that $\beta$ and $\gamma$ are linearly dependent, 
a contradiction. It follows that $n=1$ in~\eqref{eq:sumofbasisZERO},
 and hence $\alpha=c_1\basis{j_1}$, i.e. 
 $\ray=\rgb{\basis{j_1}}$.
 \end{proof}

We are now ready to show that the convex cone of loans of simple interest rate equal to 
$x$ is admissible. 

\begin{proposition}
If $x>0$ then $\sloans{x}$ is an admissible convex cone.  
\label{p:simpleinterestrateadmissible}
\end{proposition}

\begin{proof}
It suffices to apply Lemma~\ref{l:unique} and  Lemma~\ref{l:extremalray}.
\end{proof}

Proposition~\ref{p:simpleinterestrateadmissible} shows that the expression~\eqref{eq:sumofbasisZERO} 
given by Proposition~\ref{p:unique}
is the 
analysis of $\alpha$ in its extremal projections, in the sense of Definition~\ref{d:extremalprojections}.
It is useful to illustrate with an example the fact that loans in $\sloans{x}$ admit a unique expression as sum of elements 
of the extremal rays. 
\begin{example}
The loan 
$$
\alpha=-100+55\annob+60\annob^2
$$
belongs to $\sloans{0.1}$. Its analysis in its extremal projections 
is given as follows
$$
-100+55\annob+60\annob^2
=
(-50+55\annob)+(-50+60\annob^2)
$$
Indeed,  $-50+55\annob\in \elemd{0.1}{1}$
and
$-50+60\annob^2\in\elemd{0.1}{2}$. We may now apply Lemma~\ref{l:uniquely} and 
proceed to write each of the two elementary loans as the sum of a mutuum 
and an arbitrage, thus obtaining the following expression.
$$
-100+55\annob+60\annob^2
=[(-50+50\annob)
+5\annob]
+
[(-50+50\annob^2)
+
10\annob^2]
$$
This decomposition may be seen as an expression of the fact that,  
although in a non elementary loan the financial obligation of the borrower 
is split up in various parts, it still maintains a  
 \textit{unitary character}. In particular, 
 the obligation to return the principal, or part 
 of the principal, should not be disconnected from the 
 obligation to pay the corresponding interest. It seems to us that 
 the analysis of a given loan in its extremal projections 
 fits well with these general considerations.
\end{example}

\subsection{The Convex Cone Generated by Elementary Loans of 
a Given Simple Interest Rate is Arbitrage-Free}

Recall that, in our model, interest rates are subject to change every year. 

\begin{proposition}
If $x>0$ then $\sloans{x}$ is an arbitrage-free loan market. 
\label{p:af}
\end{proposition}
\begin{proof}
We have already shown that $\sloans{x}$ is an admissible convex cone. In order to show that it is arbitrage-free, it suffices to show that if   $\gamma\in\linearhull{\sloans{x}}$
then $\gamma$ is not an arbitrage. Lemma~\ref{l:linearhull} implies that $\gamma$
may be written as 
$\gamma=\alpha-\beta$, where $\alpha,\beta\in\sloans{x}$.
Let us assume that $\gamma(0)=0$. In order to show that $\gamma\not\in\arbit$ 
we have to show that either $\gamma(k)=0$ for each $k\in\NN$ or 
that there exists $k\geq 1$ such that $\gamma(k)<0$.
Observe that $\alpha(0)=\beta(0)$ and let us consider the sets 
$A\eqdef\tsupp{\alpha}\setminus\{0\}$
and
$B\eqdef\tsupp{\beta}\setminus\{0\}$. We consider the various possible cases: 
(i) $A=B$; (ii) $A\subsetneq B$; (iii) $B\subsetneq A$;  (iv)  $A\cap B=\emptyset$; 
(v) neither of the above. We will show that in each case either 
$\gamma(k)=0$ for each $k\in\NN$ or 
that there exists $k\geq 1$ such that $\gamma(k)<0$.

If (i) holds then let us consider the coefficients $c_k$ and $d_k$ which appear in the 
analysis of $\alpha$ and $\beta$, respectively, in their extremal projections. Hence 
$$
\alpha=\sum_{k=1}^{n}c_k\basis{j_k},\quad \beta=\sum_{k=1}^{n}d_k\basis{j_k}, 
\quad
c_k>0,d_k>0
$$
where $A=B=\{j_1,j_2,\ldots,j_n\}$, $n\in\NN\setminus\{0\}$.
If $c_1<d_1$ then $\alpha(j_1)<\beta(j_1)$, by~\eqref{eq:formulas}, hence 
$\gamma(j_1)<0$, and the proof is completed in this case.
If $c_k\geq d_k$ for each $k=1,2,\ldots$ then actually 
$c_k=d_k$ for each $k=1,2,\ldots$, since $\sum_k c_k=|\alpha(0)|
=|\beta(0)|=\sum_k d_k$, and then $\gamma(j_k)=0$ for each $k$, 
hence $\gamma=0$, and the proof is completed in this case.
If it is not true that 
$c_k\geq d_k$ for each $k=1,2,\ldots$ then 
let $k$ be the smallest index for which it fails. Hence 
$c_k<d_k$, and 
 then $\alpha(j_k)<\beta(j_k)$, by~\eqref{eq:formulas}, hence 
$\gamma(j_k)<0$, and the proof is completed in this case. 

If (ii) holds, write $A=\{ j_1, j_2, \ldots,j_n\}$
and
$B=\{i_1,i_2,\ldots,i_\ell\}$ with $\ell>n$. Then 
$$
\alpha=\sum_{k=1}^{n}c_k\basis{j_k},
\quad 
\beta=\sum_{k=1}^{\ell}d_k\basis{i_k}
\quad
c_k>0,d_k>0
$$
Since $A\subsetneq B$, there exists $k\in\{1,2,\ldots,\ell\}$
such that 
$i_k\not\in\{ j_1, j_2, \ldots,j_n\}$. Hence 
$$
\gamma(i_k)=\alpha(i_k)-\beta(i_k)
=0-\beta(i_k)
=-d_k\basis{i_k}(i_k)
=-d_k(1+xi_k)<0
$$
and the proof is completed in this case. 

If (iii) holds, write 
$B=\{i_1,i_2,\ldots,i_\ell\}$,
$A=
B\cup
\{ j_1, j_2, \ldots,j_n\}$
 with $1\leq n$, and 
 $\{ j_1, j_2, \ldots,j_n\}\cap B=\emptyset$.
Then 
$$
\alpha=\sum_{k=1}^{\ell}c_k\basis{i_k}+
\sum_{k=1}^{n}g_k\basis{j_k}
,
\quad 
\beta=\sum_{k=1}^{\ell}d_k\basis{i_k}
\quad
c_k>0,d_k>0,g_k>0
$$
Now observe that it is not possible that 
$d_1\leq c_1, d_2\leq c_2,\ldots,d_{\ell}\leq c_{\ell}$, since this would imply 
that 
$$
|\beta(0)|=\sum_{k=1}^{\ell}d_k
\leq
\sum_{k=1}^{\ell}c_k<
\sum_{k=1}^{\ell}c_k+
\sum_{k=1}^{n}g_k
=|\alpha(0)|
$$
in contradiction with the fact that $\alpha(0)=\gamma(0)$. 
Hence there exists $k$ such that $d_k>c_k$.
It follows that 
$$
\gamma(i_k)
=
\alpha(i_k)-\beta(i_k)
=c_k\basis{i_k}-d_k\basis{i_k}<0
$$
and the proof is completed in this case.

If (iv) holds the we may write 
\begin{equation}
\alpha=\sum_{k=1}^{n}c_k\basis{j_k},
\quad 
\beta=\sum_{k=1}^{\ell}d_k\basis{i_k}
\quad
c_k>0,d_k>0
\label{eq:v}
\end{equation}
where the sets $A=\{ j_1, j_2, \ldots,j_n\}$
and
$B=\{i_1,i_2,\ldots,i_\ell\}$
are disjoint. Hence 
$$
\gamma(i_1)=
\alpha(i_1)-\beta(i_1)=-\beta(i_1)<0
$$
and the proof is completed in this case. 

If (v) holds then we may write $\alpha$ and $\beta$ as in~\eqref{eq:v}
and there exists $k$ such that $i_k\not\in A$.
Hence 
$$
\gamma(i_k)=\alpha(i_k)-\beta(i_k)=-\beta(i_k)<0
$$
and the proof is completed in all cases. 
\end{proof}

\subsection{The Convex Cone Generated by Elementary Loans of 
a Given Simple Interest Rate is Saturated}

The following result shows that the convex cone $\sloans{x}$ is \textit{saturated}
 in a natural sense.

\begin{proposition}
If $x>0$ and $\alpha$ is a loan, then the following conditions are equivalent:
\begin{description}
\item[(a)] The collection $\sloans{x}\cup\{\alpha\}$ is arbitrage-free.  
\item[(b)] $\alpha\in\sloans{x}$.
\item[(c)] $-\alpha(0)=\sum_{k\geq1}\frac{\alpha(k)}{1+xk}$
\end{description}
\label{p:saturated}
\end{proposition}

\begin{proof}
We know from Proposition~\ref{p:af} that (\textbf{b}) implies (\textbf{a}), 
hence it suffices to show that if $\alpha\not\in\sloans{x}$ then 
the collection $\sloans{x}\cup\{\alpha\}$ is not arbitrage-free.  
We already know that (\textbf{b})  and (\textbf{c}) are equivalent, from 
Corollary~\eqref{c:corollaryintrinsic}, hence if 
$\alpha\not\in\sloans{x}$ then 
$-\alpha(0)=\sum_{k\geq1}\frac{\alpha(k)}{1+xk}$, and thus 
we may write $\alpha=\beta+w$ or 
$\alpha=\beta-w$ 
where $\beta\in\sloans{x}$ and $w\in\arbit$. In the first case, 
$w=\alpha-\beta$, thus
$w\in\linearhull{\ccgb{\sloans{x}\cup\{\alpha\}}}$, hence 
$\sloans{x}\cup\{\alpha\}$ is not arbitrage-free. In the second case, 
$w=\beta-\alpha$, and the same conclusion follows. 
\end{proof}

The following result shows that if 
a collection of economic agents is given, 
involved in the economic activies 
of buying or selling elementary 
loans at simple interest $x>0$ (with the usual understanding that 
$x$ changes from time to time in an unpredictable way) 
and a new (not elementary) loan is introduced among 
those available to the agents, then the introduction 
of this new loan will introduce arbitrages if and only if it does not belong to 
$\sloans{x}$

\begin{corollary}
If $x>0$ and $\alpha$ is a loan, then the following conditions are equivalent:
\begin{description}
\item[(a)] The collection $\elem{x}\cup\{\alpha\}$ is arbitrage-free.  
\item[(b)] $\alpha\in\sloans{x}$.
\item[(c)] $-\alpha(0)=\sum_{k\geq1}\frac{\alpha(k)}{1+xk}$
\end{description}
\end{corollary}

\begin{proof}
The proof follows at once from Proposition~\ref{p:saturated} and from the fact that 
$\sloans{x}$ is the convex cone generated by
$\elem{x}$.
\end{proof}

\section{Payoff Amounts}

The value of the payoff amounts will be computed assuming that they introduce no arbitrage
in the market. The result is the same as the one obtained in Aretusi and Mari (2018). 
We assume that the payoff amounts are established contractually at time zero, whence the requirement that no arbitrage may arise this way. 

\begin{definition}
Let $x>0$ and $\alpha\in\sloans{x}$. Let $n\geq 2$
be the maturity of $\alpha$. Assume that $\dr{\alpha}\in\polinomi$ has degree $n-1$.
Then the following conditions are equivalent:
\begin{description}
\item[(a)]
For each $k\in\{1,2,\ldots,n-1\}$,
\begin{equation}
\left[\sum_{i=0}^{k}\alpha(i)\annob^i\right]+\dr{\alpha}(k)\annob^k\in\sloans{x} 
\end{equation}

\item[(b)]
For each  $k=1,2,\ldots,n-1$, 
\begin{equation}
\dr{\alpha}(k)\eqdef
-(1+kx)\left[\alpha(0)+\frac{\alpha(1)}{1+x}
+\frac{\alpha(2)}{1+2x}
+\frac{\alpha(3)}{1+3x}
+\ldots
\frac{\alpha(k)}{1+kx}
\right]
\label{eq:debitoresiduok} 
\end{equation}
 \end{description}
The polynomial $\dr{\alpha}$ is called the polynomial \textbf{associated} 
to $\alpha$.
\end{definition}

\begin{proof}
It suffices to apply Corollary~\ref{c:corollaryintrinsic}.  
\end{proof}

Observe that (\textbf{a}) means that if only the first $k$ installments 
are payed and then, at time $k$, the amount $\dr{\alpha}(k)$ is payed, 
then this new transaction belongs to $\sloans{x}$. 
In view of Proposition~\ref{p:saturated}, 
the polynomial encodes the unique payoff amounts that 
do not introduce any arbitrage in the market.

\begin{example}
If $x=0.1$ let $\alpha=-100+55\annob+48\annob^2+26\annob^3$.
Then $\alpha\in\sloans{x}$ and 
$$
\dr{\alpha}=55\annob+24\annob^2
$$
\end{example}

\section{Main Results on The Behavior of the Payoff Amounts}

We now examine the behaviour of the
payoff amounts associated to a loan of simple interest $x>0$
whose  installments are all equal to each other.
In this section, the value of $n\geq 2$ is fixed. Since the value 
of the installments change proportionally to a change in the principal, it suffices to 
assume that the principal is equal to $1$.

\begin{definition}
If $x>0$, we define $\cil\in\loans$ as the unique 
loan with the following properties:
\begin{description}
\item[(a)] $\cil\in\sloans{x}$, $\cil$ has maturity $n$,
$\cil(0)=-1$.
\item[(b)] $\cil$ has constant installments: $\cil(1)=\cil(2)=\ldots=\cil(n)$. 
\end{description}
\end{definition}
Thus, $\cil$ encodes a loan at simple interest equal to $x$ where 
the principal is equal to $1$ and the installments are constant.

\begin{lemma}
The constant installements of $\cil$ are equal to 
$\vcil$ where
$$
\vcil\eqdef{\left(\sum_{k=1}^n\frac{1}{1+xk}\right)}^{-1}
$$
\end{lemma}

\begin{proof}
It suffices to observe that if $r$ is the common value of $\cil(k)$ then 
$$
1=\sum_{k=1}^n\frac{r}{1+xk}=r\sum_{k=1}^n\frac{1}{1+xk}
$$
\end{proof}

\paragraph{Notation.} It is convenient to adopt the following notation.
$$
w(x,n)\eqdef\frac{1}{\vcil}=\sum_{k=1}^n\frac{1}{1+xk}
$$

Recall that $\drcil\in\polinomi$
encodes the payoff amounts associated to $\cil$. We are interested in the  behaviour of the coefficients of $\drcil$.

\begin{definition}
If $n\geq 2$ we define $a_n$ as the only positive solution of the following equation.
\begin{equation}
1=x\cdot\left(\sum_{k=1}^n\frac{1}{1+kx}\right)
\label{eq:criticalvalue} 
\end{equation}
The number $a_n$ will be denoted \textbf{critical rate}.
\end{definition}
Observe that the function on the right-hand side of~\eqref{eq:criticalvalue}
is strictly increasing for $x>0$, its value at $0$ is $0$, and its limiting value 
as $x\to+\infty$ is strictly greater than 1. Hence there is one and only one 
positive solution to~\eqref{eq:criticalvalue}.
The following result follows at once from these observations.
\begin{lemma}
If $x>0$ then $x<a_n$ if and only if $xw(x,n)<1$.
\label{l:function} 
\end{lemma}
 
\begin{example}
Observe that $a_2$ is the golden ratio
$$
a_2=\frac{1+\sqrt{5}}{2}
$$
\end{example}

\begin{lemma} The critical rates form a decreasing sequence:
If $n=2,3,\ldots$ then
$$
a_2>a_{3}>a_4>\ldots
$$ 
\end{lemma}

\begin{proof}
Observe that if $x>0$ 
then the function $w(x,n)$ is strictly increasing as a function of $n$. 
Since $a_{n+1}w(a_{n+1},n+1)=1$ and $a_{n+1}>0$ 
it follows that 
$a_{n+1}w(a_{n+1},n)<a_{n+1}w(a_{n+1},n+1)=1$ hence 
$a_{n+1}w(a_{n+1},n)<1$ and now Lemma~\ref{l:function} implies that 
$a_{n+1}<a_{n}$. 
\end{proof}

\begin{theorem}
If $x>0$ then the following conditions are equivalent:
\begin{description}
\item[(a)] $x<a_n$
\item[(b)] $\drcil(1)<1$
\end{description}
\label{thm:lessthan1}
\end{theorem}
Observe that (\textbf{b}) is what we would like to expect from 
the payoff amounts. However, in loans at simple interest this is not necessarily the case.  

\begin{theorem}
If $x>0$ then the following conditions are equivalent:
\begin{description}
\item[(a)] $x>a_n$
\item[(b)] $\drcil(1)>1$
\end{description}
\label{thm:larger}
\end{theorem}

\begin{lemma}
If $1>\drcil(1)$ then 
$\drcil(1)>\drcil(2)$. 
\label{l:firstcase}
\end{lemma}

The following result is an extension of Lemma~\ref{l:firstcase}.

\begin{proposition}
If  $\drcil(k)>\drcil(k+1)$ for some $k$
then 
$\drcil(k+1)>\drcil(k+2)$.
\label{p:mainlemma}
\end{proposition}

The following result is an immediate corollary of the previous results. 

\begin{corollary}
If $x<a_n$ then the sequence
${\{\drcil(k)\}}_k$ is decreasing. 
If $x>a_n$ then the sequence
${\{\drcil(k)\}}_k$ will be increasing at first 
but once it starts decreasing it will continue to do so.
\end{corollary}

\section{Proofs}

\paragraph{Proof of Theorem~\ref{thm:lessthan1}}
\begin{proof}
Define 
$$
w(x,n)\eqdef\frac{1}{\vcil}
$$
and observe that 
$$
\drcil(1)=\left(1-\frac{\vcil}{1+x}\right)(1+x)=
1+x-\vcil
$$
Hence $\drcil(1)<1$ if and only if 
$x<\vcil$ which is equivalent to 
$xw(x,n)<1$. The conclusion follows from the definition of the critical rate. 
\end{proof}

\paragraph{Proof of Theorem~\ref{thm:larger}}

\begin{proof}
Observe as before that 
$$
\drcil(1)=\left(1-\frac{\vcil}{1+x}\right)(1+x)=
1+x-\vcil
$$
Hence $\drcil(1)>1$ if and only if 
$x>\vcil$ which is equivalent to 
$xw(x,n)>1$. The conclusion follows from the definition of the critical rate. 
\end{proof}

\paragraph{Proof of Lemma~\ref{l:firstcase}}

\begin{proof}
Let us assume that $1>\drcil(1)$. We will show that 
$\drcil(1)>\drcil(2)$. 
Observe that 
$$
\drcil(2)=\left(1-\frac{\vcil}{1+x}-\frac{\vcil}{1+2x}\right)(1+2x)
=\left(1-\frac{1}{(1+x)w(x,n)}-\frac{1}{(1+2x)w(x,n)}\right)(1+2x)
$$
and recall that 
$$
\drcil(1)=1+x-\vcil
$$
hence
$\drcil(2)<\drcil(1)$ if and only if 
$$
(1+2x)-{\frac{1+2x}{w(n,x)(1+x)}}-{\frac{1+2x}{w(n,x)(1+2x)}}<1+x-\frac{1}{w(n,x)}
$$
which is equivalent to 
$$
1+2x-{\frac{1+2x}{w(n,x)(1+x)}}-{\frac{1}{w(n,x)}}<1+x-\frac{1}{w(n,x)}
$$
which in turn is equivalent to 
\begin{equation}
x-{\frac{1+2x}{w(n,x)(1+x)}}<0
\label{eq:bridge} 
\end{equation}
Now observe that 
$$
{\frac{1}{w(n,x)}}<{\frac{1+2x}{w(n,x)(1+x)}}
$$
and recall that we know, from Theorem~\ref{thm:lessthan1}, that 
$$
x<{\frac{1}{w(n,x)}}
$$
Hence~\eqref{eq:bridge} follows at once. 
\end{proof}

We are now ready to show that the content of Lemma~\ref{l:firstcase}
holds in general.

\paragraph{Proof of Proposition~\ref{p:mainlemma}}
\begin{proof}
The proof is similar to that of Lemma~\ref{l:firstcase}.
Observe that 
$$
\drcil(k)=\left[
1-\sum_{i=1}^{k}\frac{\vcil}{1+ix}
\right]
(1+kx)
=1+kx-\vcil\left[
\sum_{i=1}^{k}\frac{1+kx}{1+ix}
\right]
$$ 
and 
$$
\drcil(k+1)=\left[
1-\sum_{i=1}^{k+1}\frac{\vcil}{1+ix}
\right]
(1+kx+x)
=1+kx+x-\vcil\left[
\sum_{i=1}^{k+1}\frac{1+kx+x}{1+ix}
\right]
$$ 
Hence the condition that 
$\drcil(k)>\drcil(k+1)$ means that 
$$
x-\vcil\left[
\sum_{i=1}^{k+1}\frac{1+kx+x}{1+ix}
\right]<-\vcil\left[
\sum_{i=1}^{k}\frac{1+kx}{1+ix}
\right]
$$
which is equivalent to 
\begin{equation}
x-\vcil\left[
\sum_{i=1}^{k}\frac{1+kx+x}{1+ix}
\right]<-\vcil\left[
\sum_{i=1}^{k-1}\frac{1+kx}{1+ix}
\right] 
\label{eq:generalcase}
\end{equation}
and since 
$$
x-\vcil\left[
\sum_{i=1}^{k}\frac{1+kx+x}{1+ix}
\right]
=
x-\vcil\left[
\sum_{i=1}^{k-1}
\frac{1+kx}{1+ix}
+
\frac{1+kx}{1+kx}
+
\sum_{i=1}^{k}
\frac{x}{1+ix}
\right]
$$
it follows that~\eqref{eq:generalcase} is equivalent to 
\begin{equation}
x-\vcil\left[
1+
\sum_{i=1}^{k}
\frac{x}{1+ix}
\right]
<0
\label{eq:generalcasesimplified} 
\end{equation}
Now observe that 
$$
\drcil(k+2)
=1+kx+2x-\vcil\left[
\sum_{i=1}^{k+2}\frac{1+kx+2x}{1+ix}
\right]
$$ 
and in order to show that 
$\drcil(k+1)>\drcil(k+2)$ it is necessary and sufficient to show that 
$$
1+kx+2x-\vcil\left[
\sum_{i=1}^{k+2}\frac{1+kx+2x}{1+ix}
\right]
<1+kx+x-\vcil\left[
\sum_{i=1}^{k+1}\frac{1+kx+x}{1+ix}
\right]
$$
which is equivalent to 
\begin{equation}
x-\vcil\left[
\sum_{i=1}^{k+2}\frac{1+kx+2x}{1+ix}
\right]
<-\vcil\left[
\sum_{i=1}^{k+1}\frac{1+kx+x}{1+ix}
\right]
\label{eq:toshow} 
\end{equation}
Since we know that~\eqref{eq:generalcasesimplified}
holds, we have 
$$
x-\vcil\left[
\sum_{i=1}^{k+2}\frac{1+kx+2x}{1+ix}
\right]
<
\vcil\left[
1+
\sum_{i=1}^{k}
\frac{x}{1+ix}
\right]
-\vcil\left[
\sum_{i=1}^{k+2}\frac{1+kx+2x}{1+ix}
\right]
$$
It follows that in order to show that~\eqref{eq:toshow} holds it is sufficient to show that 
$$
\vcil\left[
1+
\sum_{i=1}^{k}
\frac{x}{1+ix}
\right]
-\vcil\left[
\sum_{i=1}^{k+2}\frac{1+kx+2x}{1+ix}
\right]
<-\vcil\left[
\sum_{i=1}^{k+1}\frac{1+kx+x}{1+ix}
\right]
$$
which is equivalent to 
$$
\left[
1+
\sum_{i=1}^{k}
\frac{x}{1+ix}
\right]
-\left[
\sum_{i=1}^{k+2}\frac{1+kx+2x}{1+ix}
\right]
<-\left[
\sum_{i=1}^{k+1}\frac{1+kx+x}{1+ix}
\right]
$$
which in turn is equivalent to 
$$
\sum_{i=1}^{k}
\frac{x}{1+ix}
-
\sum_{i=1}^{k+1}\frac{1+kx+2x}{1+ix}
<-
\sum_{i=1}^{k+1}\frac{1+kx+x}{1+ix}
$$
and this is equivalent to 
$$
\sum_{i=1}^{k}
\frac{x}{1+ix}
-
\sum_{i=1}^{k+1}\frac{2x}{1+ix}
<
-\frac{x}{1+kx+x}
-
\sum_{i=1}^{k}\frac{x}{1+ix}
$$
which is equivalent to 
$$
\sum_{i=1}^{k}
\frac{x}{1+ix}
-\frac{2x}{1+kx+x}
-\sum_{i=1}^{k}\frac{2x}{1+ix}
<
-\frac{x}{1+kx+x}
-
\sum_{i=1}^{k}\frac{x}{1+ix}
$$
which in turn is equivalent to 
$$
-\sum_{i=1}^{k}
\frac{x}{1+ix}
-\frac{2x}{1+kx+x}
<
-\frac{x}{1+kx+x}
-
\sum_{i=1}^{k}\frac{x}{1+ix}
$$
and this is finally equivalent to
$$
-\frac{2x}{1+kx+x}
<
-\frac{x}{1+kx+x}
$$
and this last condition holds for each $x>0$
\end{proof}

\section*{References}

Aretusi, G., Mari, C.: 
Sull’esistenza e unicità dell’ammortamento dei prestiti in regime lineare. 
Il Risparmio.
Anno LXVI – n. 1 gennaio - luglio 2018

\smallskip
\noindent
Cacciafesta, F.: Lezioni di Matematica Finanziaria Classica e Moderna.
Giappichelli, Torino (2001)

\smallskip
\noindent
CFPB (Consumer Financial Protection Bureau): What is a payoff amount? Is my payoff amount the same as my current balance?
\url{https://www.consumerfinance.gov/ask-cfpb/what-is-a-payoff-amount-is-my-payoff-amount-the-same-as-my-current-balance-en-205/} (2020). Accessed 26 June 2023

\smallskip
\noindent
Ingrao, B., Israel, G.: La mano invisibile: l'equilibrio economico nella storia della scienza.
Laterza, Bari (1996)

\smallskip
\noindent
Levi, E.: Corso di Matematica Finanziaria. La Goliardica, Milano (1953)

\smallskip
\noindent
Schneider, R.: Convex cones: Geometry and Probability. Lecture Notes in Mathematics 2319. Springer, Berlin (2022)

\smallskip
\noindent
Varoli, G.: Matematica Finanziaria. Pàtron Editore, Bologna (1983)

\bigskip

\textit{E-mail addresses:} 

F. di Biase:\,\texttt{fausto.dibiase@unich.it} (corresponding author)

S. Di Rocco:\,\texttt{stefano.dirocco@studenti.unich.it}

A. Ortolano:\,\texttt{alessandra.ortolano@unitus.it}

M. Parton:\,\texttt{maurizio.parton@unich.it}

\smallskip

\textit{Orcid:} 

F. Di Biase: \texttt{0000-0003-0294-6919}

A. Ortolano: \texttt{0000-0001-8989-0780}

M. Parton: \texttt{0000-0003-4905-3544}

\end{document}